\documentclass{jfm}
\usepackage{graphicx}
\usepackage{natbib}
\usepackage{color}
\usepackage{enumitem}
\usepackage{latexsym}
\usepackage{subfig}
\usepackage{natbib}
\usepackage{upgreek}
\usepackage{mathrsfs}
\usepackage{float}
\usepackage{caption}
\usepackage{soul}
\usepackage{textcomp}
\usepackage{stmaryrd}
\usepackage{amsmath}
\usepackage{mathtools}
\usepackage{amsfonts}
\usepackage{epstopdf, epsfig}
\usepackage{amsmath,etoolbox}
\usepackage{bigints}
\usepackage{lipsum}
\usepackage{physics}
\usepackage[normalem]{ulem}

\usepackage{xargs}                      
\usepackage[pdftex,dvipsnames]{xcolor}  
\usepackage[colorinlistoftodos,prependcaption,textsize=small]{todonotes}

\AtBeginEnvironment{pmatrix}{\setlength{\arraycolsep}{2pt}}

\def\ee{{\rm e}}
\def\ii{{\rm i}}

\title{Energy transfer in resonant and near-resonant internal wave triads for weakly non-uniform stratifications. Part I: Unbounded domain}

\author{G. Saranraj\aff{1} \and Anirban Guha\aff{1}
  \corresp{\email{anirbanguha.ubc@gmail.com}}
 }

\affiliation{
\aff{1} School of Science and Engineering, University of Dundee, DD1 4HN, U.K.
}

\begin{document}

\maketitle
\begin{abstract}
In this paper, using multiple scale analysis we derive a generalized mathematical model for amplitude evolution, and for calculating the energy exchange in resonant and near-resonant {global} triads consisting of weakly nonlinear internal gravity wave packets in weakly non-uniform density stratifications in an unbounded domain in the presence of viscous and rotational effects. Such triad interactions are one of the mechanisms by which high wavenumber internal waves lead to ocean turbulence and mixing via parametric subharmonic instability.  Non-uniform stratification introduces detuning -- mismatch in the vertical wavenumber triad condition, which may strongly affect the energy transfer process. We investigate in detail how factors like wave-packets' width, group speeds, nonlinear coupling coefficients, detuning, and viscosity affect energy transfer in weakly varying stratification. We also investigate the effect of detuning on energy transfer in varying stratification for different daughter wave combinations of a fixed parent wave. We find limitations of the well-known `pump-wave approximation' and derive a non-dimensional number, which can be evaluated from initial conditions, that can predict the maximum energy transferred from the parent wave during the later stages. Two additional non-dimensional numbers, based on various factors affecting energy transfer  between near-resonant wave-packets have also been defined. 
Moreover, we identify the optimal background stratification in a medium of varying stratification for the parent wave to form a triad with no detuning so that the energy transfer is maximum.

\end{abstract}

\begin{keywords} 
Internal gravity waves, wave triads, nonlinear density stratification, parametric subharmonic instability
\end{keywords}

\section{Introduction}
 Internal gravity waves are often produced in  oceans when the stably stratified ocean water is driven back and forth over submarine topography by tidal currents.    
Low-mode internal gravity waves have long wavelengths, and can travel long distances from their generation site  without dissipation \citep{laurent}. Understanding the mechanism(s) behind the breakdown of these waves is an active area of research, since it  finally leads to small scale ocean mixing. One of the plausible mechanisms through which this breakdown occurs is  parametric subharmonic instability (PSI) --  a nonlinear interaction between waves forming a resonant triad by which energy is transferred from low wavenumber, high
frequency modes to high wavenumber, low frequency modes \citep{Mac_2005}.
 In a resonant internal gravity wave triad, a primary (or parent) wave of angular frequency $\omega_{3}$ and wavevector $\mathbf{k}_{3}$  resonantly forces two daughter waves (indexed by `$1$' and `$2$') by transferring its own energy, when both the conditions $\omega_{3} = \omega_{1} + \omega_{2}$ and $\mathbf{k}_{3} = \mathbf{k}_{1} + \mathbf{k}_{2}$ are met \citep{hasselmann_1967}. 
For a given parent wave, it is possible to have infinite daughter waves satisfying the resonant triad condition.

From laboratory experiments and theoretical analyses, \cite{bourget} showed that in a uniformly stratified fluid, the growth rate of the daughter waves depends on the wavenumber, frequency and Reynolds number of the parent wave. Since ocean's density stratification is non-uniform, recent efforts have been directed towards understanding energy transfer in non-uniformly stratified fluids. Triads in a non-uniform stratification behave differently because of the wavenumbers' dependence on the stratification. Monochromatic internal gravity waves are an exact solution to the fully nonlinear Navier-Stokes equation in a uniformly stratified fluid \citep{lighthill2001waves}. The same is not true when the fluid is non-uniformly stratified; moreover, a given mode can interact with itself. Through such self interaction, a primary mode in a non-uniform stratification can yield superharmonic daughter modes having twice of the primary's horizontal wavenumber {and angular frequency} \citep{sutherland}. 
 However, \cite{sutherland} did not find any occurrence of PSI.
 \cite{diamessis} showed that superharmonics mainly form when the pycnocline is sharp. Similar conclusions were obtained in \cite{gayen}; they showed that the energy transfer through PSI is negligible when the parent waves have vertical wavelength comparable to the pycnocline thickness. However, significant energy transfer through PSI is observed when the vertical wavelength of the waves are nearly an order of magnitude lesser than the pycnocline thickness. 
Using a weakly nonlinear analysis, \cite{wunsch} studied the self interaction of a low mode internal gravity wave assuming the stratification to be layerwise constant, and found that self-interaction of a primary mode can resonantly force superharmonic waves, similar to what was concluded in \cite{sutherland}.
\cite{varma} provided the necessary conditions for a mode to resonantly force other modes (through self interaction or by interaction with other modes) in a general non-uniform stratification using weakly nonlinear analysis. From these previous studies, it can be inferred that the length scale of stratification plays a key role in determining the cascading process of the primary mode, that is, whether it will be superharmonic or subharmonic.

Higher modes are far less studied, they can lead to small scale turbulence and mixing via PSI type triad interactions \citep{laurent}.  
Energy flux estimation in the Mid-Atlantic Ridge has revealed that high modes (e.g.\ modes $10$--$25$) contain about 18$\%$ of the total flux \citep{laurent,laurent_2003}. Additionally, internal wave beams having higher modes are also not uncommon in oceans. For example, $\mathrm{M_2}$ internal gravity wave beams composed of high wavenumbers (expected to more than mode 100) have been observed in the seismic images of the Norwegian sea \citep{holbrook}. {Moreover, a recent study combining semi-analytical model with observations (satellite and \emph{in-situ} measurements) has revealed that high modes (modes $>10$) account for a relatively large fraction (27$\%$) of total tidal energy conversion in the oceans \citep{vic2019deep}.}  

\begin{figure}
{\centering
\includegraphics[width=1.0\linewidth,keepaspectratio]{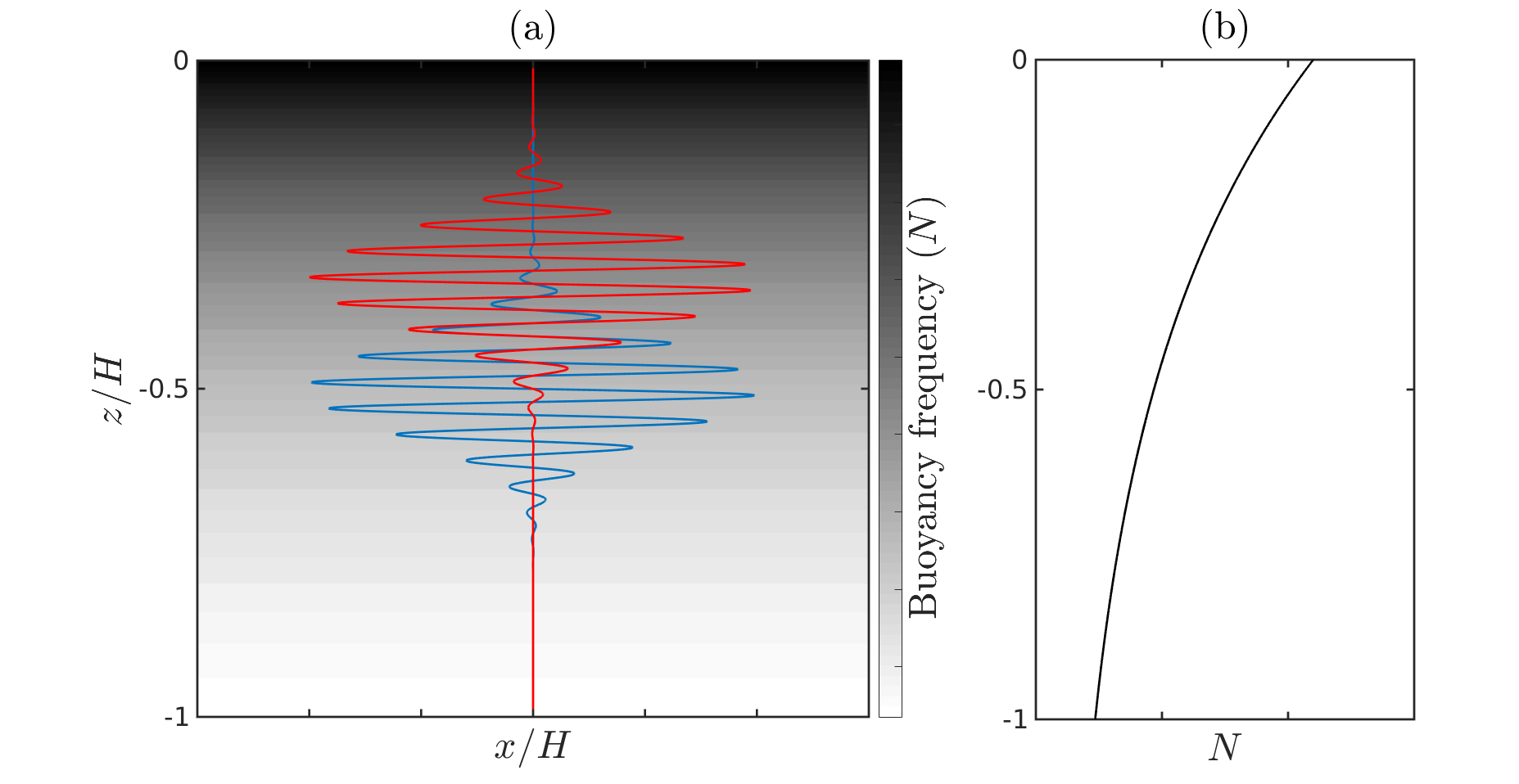}}
\caption[Energy evolution plot for triad case 4]{{ (a) Wave-packets interacting in a medium of varying stratification. (b) The buoyancy frequency ($N$) profile used in (a) is {similar to the profile used in \cite{laurent_2003}}. }}
\label{fig:schematic}
\end{figure}

In this paper, we have focused on internal wave triads whose constituent waves have vertical wavelengths at least an order of magnitude lesser than the length scale of buoyancy frequency's variation in the $z$-direction. {A simple schematic of such wave-packets interacting  in a weakly varying stratification is shown in figure \ref{fig:schematic}}. Such buoyancy frequency profiles in deep ocean stratification are common, and have been considered in \cite{laurent}, \cite{munk_strat} and  \cite{zhao_strat}.
In such slowly varying stratification profiles, the vertical wavenumber of the higher mode internal waves undergoes a slow variation in space as they propagate vertically, unlike what happens in rapidly varying stratifications.
In addition to resonant triads, we have also focused on near-resonant triads, that is, waves which \emph{almost} satisfy the triad condition. Such triads have previously been studied by \cite{lambnear}; it was shown that near-resonant triads can occur when internal gravity waves generated via tide--topography interactions interact among themselves. The interaction strength was also found to be comparable to that of an exact triad. 

The paper is organized as follows. In \S \ref{Section:2}, we derive a significantly general amplitude evolution equations of the constituent waves of a resonant and a near-resonant triad.
To obtain these equations, we have reduced the viscous, incompressible, two-dimensional (2D) Boussinesq Navier-Stokes equations in the $f$-plane by assuming the streamfunction, $y-$direction velocity, and the corresponding buoyancy perturbation due to each wave to be a product of slowly varying amplitude and rapidly varying phase. In \S \ref{Section:3}, we use normal mode analysis to study  triad interaction in uniform stratification, and also focus on the limitations of using normal modes. Additionally in \S \ref{Section:4}, we analyze the energy transfer between near-resonant finite width wave-packets in uniform stratification. 
In \S \ref{Section:5}, we study the factors affecting the energy transfer between near-resonant finite width wave-packets in varying stratification using the equations derived in \S \ref{Section:2}.
In \S \ref{section:5.1} the various factors which effect the energy transfer between inviscid wave-packets in varying stratification are investigated.
In \S \ref{section:5.2} the effects of viscosity on the growth rates of daughter waves in varying stratification are analyzed, and an expression for the normal mode growth rate is also derived. In \S \ref{sec:optimal_base}, we estimate the optimal base stratification that transfers maximum energy in a varying stratification. 
The results obtained from  multiple scale analysis are numerically validated in \S \ref{section:5.3}.  
{{In \S \ref{Section:5.5}, we show that mismatch in vertical wavenumber, for a given change in stratification, can be an important factor in deciding how much energy a particular daughter wave combination can extract from the parent wave.}} 
The paper is summarized and concluded in \S \ref{Section:7}.


\section{Derivation of the governing equations \label{Section:2}}

The {viscous}, incompressible, 2D (in the $x$--$z$ plane) Boussinesq Navier-Stokes equations {in the $f$-plane}, in the absence of a background flow, can be compactly written in terms of the perturbation streamfunction $\psi$, the perturbation buoyancy $b$, {and the velocity along $y$-direction $v$}, as follows:
\renewcommand{\theequation}{\arabic{section}.\arabic{equation}a}
\begin{equation}
\frac{\partial }{\partial t}\left(\nabla^{2}\psi\right)  = -\{\nabla^{2}\psi,\psi\} - \frac{\partial b}{\partial x} + f\frac{\partial v}{\partial z} + \nu\Delta^{2}\psi, \label{eqn:NS_stream}
\end{equation}
\addtocounter{equation}{-1}
\renewcommand{\theequation}{\arabic{section}.\arabic{equation}b}
\begin{equation}
\frac{\partial v}{\partial t} + f\frac{\partial \psi}{\partial z}  = -\{v,\psi\} + \nu\nabla^{2}v. \label{eqn:corilios}
\end{equation}
\addtocounter{equation}{-1}
\renewcommand{\theequation}{\arabic{section}.\arabic{equation}c}
\begin{equation}
\frac{\partial b}{\partial t} - N^{2}(\epsilon_{n} z)\frac{\partial \psi}{\partial x}  = -\{b,\psi\}. \label{eqn:material_cons}
\end{equation}
\renewcommand{\theequation}{\arabic{section}.\arabic{equation}}

\noindent Here $N^{2} \equiv -\left(g/\rho^{*}\right)\left(d \bar{\rho}/d z\right)$  is the squared buoyancy frequency,  $\bar{\rho}$ is the base density profile and  $\rho^{*}$ is the reference density, $g$ is the acceleration due to gravity, {$f$ is the Coriolis frequency, and $\nu$ is the kinematic viscosity.} The perturbation buoyancy is defined as $b \equiv -g\rho/\rho^{*}$, where $\rho$ is the perturbation density. The buoyancy frequency is assumed to vary \emph{weakly} with $z$, the parameter $\epsilon_{n}$, provides a quantitative measure of this weak variation.
 The Poisson bracket is defined as $\{\mathfrak{B}_1,\mathfrak{B}_2\} \equiv  (\partial \mathfrak{B}_1/\partial x)(\partial \mathfrak{B}_2/\partial z) - (\partial \mathfrak{B}_1/\partial z)(\partial \mathfrak{B}_2/\partial x)$. 
 The squared delta operator is defined as $\Delta^2 \equiv \partial^{4}/\partial x^4 + 2\partial^{4}/\partial z^2 \partial x^2 + \partial^{4}/\partial z^4$.
 

Instead of solving the fully nonlinear equations (\ref{eqn:NS_stream})--(\ref{eqn:material_cons}) numerically, we combine (\ref{eqn:NS_stream})--(\ref{eqn:material_cons}) into a single equation and employ a multiple scale analysis. In this regard we perform $\partial( \ref{eqn:NS_stream})/\partial t  - \partial(\ref{eqn:material_cons})/\partial x + f\partial(\ref{eqn:corilios})/\partial z$, which results in
\begin{align}
\frac{\partial^{2} }{\partial t^{2}}\left(\nabla^{2}\psi\right) + N^{2}(\epsilon_{n} z)\frac{\partial^{2} \psi}{\partial x^{2}} + f^{2}\frac{\partial^{2} \psi}{\partial z^{2}} = &-\frac{\partial}{\partial t} \left(\{\nabla^{2}\psi,\psi\}\right) + \frac{\partial }{\partial x}\left(\{b,\psi\}\right) - f\frac{\partial }{\partial z}\left(\{v,\psi\}\right) \nonumber \\
&+ \nu\frac{\partial }{\partial t}\left(\Delta^{2}\psi\right) + \nu f\frac{\partial}{\partial z}\left(\nabla^{2}v\right). 
\label{eqn:combined}
\end{align}
For performing multiple scale analysis, we assume wavelike perturbations, and the streamfunction due to the $j$-th  wave ($j = 1,2,3$ since we will be considering a wave-triad) is given according to the following ansatz:
\begin{equation}
\psi_{j} = a_{j}(\epsilon_{x} x,\epsilon_{z} z,\epsilon_{t} t)F_{j}(z) \ee^{\ii(k_{j}x-\omega_{j}t)}  + \mathrm{c.c}., 
\label{eqn:stream_ans}
\end{equation}
\noindent where `c.c' denotes the complex conjugate, $a_{j}$ is the slowly varying complex amplitude, $k_j$ is the horizontal wavenumber and $\omega_j$ is the angular frequency of the  $j$-th  wave, and $F_j(z)$ is the vertical structure of a $j$-th wave.  {Similar to $\epsilon_{n}$, small parameters $\epsilon_{t}$, $\epsilon_{x}$  and  $\epsilon_z$ are respectively used to denote the weak variation of the amplitude function with time, streamwise ($x$)  and vertical ($z$) directions. Moreover, a small parameter $\epsilon_a$ signifies the order of magnitude of a wave's streamfunction amplitude. Scaling analysis to find the relations between these small parameters is given in Appendix \ref{app:A}.}

The buoyancy perturbation, corresponding to the streamfunction assumed in (\ref{eqn:stream_ans}), at the leading order $(\mathcal{O}(\epsilon_{a}))$ is given by:
\begin{equation}
b_{j} = -\frac{N^{2}(\epsilon_{n} z) k_{j}}{\omega_{j}} a_{j}(\epsilon_{x} x,\epsilon_{z} z,\epsilon_{t} t)F_{j}(z) \ee^{\ii(k_{j}x-\omega_{j}t)} + \mathrm{c.c}. 
\label{eqn:buo_ans}
\end{equation}

{The above expression is obtained via polarization relation, i.e. by substituting streamfunction expression  (\ref{eqn:stream_ans}) in (\ref{eqn:material_cons}), see  \cite{sutherland_2010,bourget}}.


{ The $y$-direction velocity is given by:}
\begin{equation}
v_{j} = -\frac{\ii f}{\omega_{j}} a_{j}(\epsilon_{x} x,\epsilon_{z} z,\epsilon_{t} t)\frac{d F_{j}}{dz} \ee^{\ii(k_{j}x-\omega_{j}t)} + \mathrm{c.c}. 
\label{eqn:CV_ans}
\end{equation}
The streamfunction (\ref{eqn:stream_ans}), the buoyancy perturbation (\ref{eqn:buo_ans}) and  the $y-$direction velocity (\ref{eqn:CV_ans})  ansatzes are substituted in (\ref{eqn:combined}). At leading order $(\mathcal{O}(\epsilon_{a}))$, the governing equation reduces to an eigenvalue problem
\begin{equation}
\frac{d^{2} F_{j}}{d z^{2}} + k_j^{2}\left(\frac{N^{2}(\epsilon_{n} z)-\omega_j^{2}}{\omega_j^{2}-f^2}\right)F_{j} = 0,
\label{eqn:eigen}
\end{equation}
solving which we can obtain the vertical structure $F_j(z)$ of the $j$-th  wave.  
For weakly varying stratification, we can use the Wentzel--Kramers--Brillouin (WKB) method and solve  (\ref{eqn:eigen}). The solution for $F_j$ up to the second order accuracy, is given by
\begin{equation}
F_{j} = \dfrac{1}{\sqrt{\abs{m_{j}}}}{\exp(\ii \int_{-\infty}^{z} m_{j} dz )}, 
\label{eqn:vert_wkb}
\end{equation}
where 
\begin{equation}
    m_{j}(\epsilon_{n} z) \equiv \pm k_{j}\sqrt{\frac{N^{2}(\epsilon_{n} z)-\omega_j^{2}}{\omega_j^{2}-f^2}}
    \label{eqn:vertical_wavenumber}
\end{equation}
is  the vertical wavenumber. { We mention in passing that the unit of amplitude function $a_j$ is $\textnormal{m}^{3/2}\textnormal{s}^{-1}$ because of the form of the streamfunction assumed.} 


The small parameter $\epsilon_n$ is included in the argument of the buoyancy frequency to emphasize that the buoyancy frequency is a function of $\epsilon_n z$, and not $z$.
The quantity $\epsilon_{z}$ is decided as follows: 
\begin{equation}
\epsilon_{z} =  \textnormal{max} \left(\epsilon_n\hspace{0.1cm}, \hspace{0.1cm}  \abs{\frac{\Delta m}{m_3}}  \right),
\label{eqn:eps_z}
\end{equation} 
{where $\Delta m \equiv m_{3}-m_{1}-m_{2}$ is the vertical wavenumber mismatch at any location in space. We have used separate small parameters for the variation of amplitude in the $z$--direction and the buoyancy frequency since they can in general be independent of each other. For example, near-resonant triads with vertical wavenumber mismatch can occur even in a uniform stratification ($\epsilon_{n} = 0$), but the amplitude of the waves will still vary in space ($\epsilon_z \neq 0$).}    
At the leading order $(\mathcal{O}(\epsilon_{a}))$, the waves satisfy the dispersion relation and behaves as a linear wave. However, at $\mathcal{O}(\epsilon^2)$ (that is, terms such as $\mathcal{O}(\epsilon_{a}\epsilon_{z}), \mathcal{O}(\epsilon_{a}\epsilon_{t}), \mathcal{O}(\epsilon_{a}\epsilon_{x}), \mathcal{O}(\epsilon_{a}^{2}))$,  triad interactions (through the nonlinear terms) slowly modulate the amplitude of each constituent wave. {We have considered the effect of viscosity at $\mathcal{O}(\epsilon^2)$, following the approach of  \cite{karimi_2014} and \cite{Near_aky}.} To study the triad interactions between the waves, the $\mathcal{O}(\epsilon^2)$ terms are gathered after substituting the streamfunction (\ref{eqn:stream_ans}), the $y$ direction velocity (\ref{eqn:CV_ans})  and the buoyancy perturbation (\ref{eqn:buo_ans}) ansatzes in  (\ref{eqn:combined}). For convenience, we denote the phase part by $\mathfrak{P}_{j}$,
\begin{equation*}
        \mathfrak{P}_{j} \equiv F_{j}(z)\ee^{\ii(k_{j}x-\omega_{j}t)}. 
\end{equation*}

The LHS is then given by:

 \begin{align}
 \noindent \textnormal{LHS} &= \sum_{j=1}^{3} \hspace{0.1cm} \frac{\partial^{2} \left[\nabla^{2}( a_{j} \mathfrak{P}_{j})\right]}{\partial t^{2}} + N^{2}(\epsilon_{n} z)\frac{\partial^{2} (a_{j} \mathfrak{P}_{j})}{\partial x^{2}} + f^{2}\frac{\partial^{2} (a_{j} \mathfrak{P}_{j})}{\partial z^{2}} + \mathrm{c.c}. \nonumber \\
&=\sum_{j=1}^{3} \hspace{0.1cm} 2\ii\left[(k^{2}_{j}+m^{2}_{j})\omega_{j}\frac{\partial a_{j}}{\partial  t}- m_{j}(\omega^{2}_{j}-f^2) \frac{\partial a_{j}}{\partial z} +  k_{j} (N^{2}(\epsilon_{n} z)-\omega^{2}_{j}) \frac{\partial a_{j}}{\partial x} \right]\mathfrak{P}_{j} \nonumber \\
&\hspace{0.87cm}+ \ii \left[ \nu a_j(k_j^2+m_j^2)\left(\omega_j(k_j^2+m_j^2) + \frac{f^2 m_j^{2}}{\omega_j} \right) \right]\mathfrak{P}_{j}
    + \mathrm{c.c}.  
 \label{eqn:LHS_triad}
 \end{align}


\noindent In \eqref{eqn:LHS_triad}, the $\mathcal{O}(\epsilon^2)$ terms are obtained when any differential operator acts on $a_j$ exactly once, e.g.\, $\partial a_j/\partial t \sim \mathcal{O} (\epsilon_a \epsilon_t)$. Next we consider the nonlinear terms.
The first term of the RHS is 
\begin{align}
\frac{\partial \left(\{\nabla^{2}\psi,\psi\}\right)}{\partial t} = & \hspace{0.4cm} \ii(\omega_{1}+\omega_{2})a_{1}a_{2}\left[(k_{1}m_{2}-k_{2}m_{1})\left(m^{2}_{2}+k^{2}_{2}-k^{2}_{1}-m^{2}_{1}\right) \right]\mathfrak{P}_{1}\mathfrak{P}_{2} \nonumber \\ 
&+\ii(\omega_{3}-\omega_{2})a_{3}\bar{a}_{2}\left[(k_{3}m_{2}-k_{2}m_{3})\left(m^{2}_{3}+k^{2}_{3}-k^{2}_{2}-m^{2}_{2}\right) \right]\mathfrak{P}_{3}\bar{\mathfrak{P}}_{2} \nonumber \\
&+\ii(\omega_{3}-\omega_{1})a_{3}\bar{a}_{1}\left[(k_{3}m_{1}-k_{1}m_{3})\left(m^{2}_{3}+k^{2}_{3}-k^{2}_{1}-m^{2}_{1}\right) \right]\mathfrak{P}_{3}\bar{\mathfrak{P}}_{1}  + \mathrm{c.c}.  \label{eqn:NL_term1} ,
\end{align}
while the second term is given by
\begin{align}
\frac{\partial \left(\{b,\psi\}\right)}{\partial x} = &\ii N^{2}(k_{1}+k_{2})a_{1}a_{2}\left[
\left(\frac{k_1}{\omega_1}-\frac{k_2}{\omega_2}\right)(k_1m_2-k_2m_1) \right] \mathfrak{P}_{1}\mathfrak{P}_{2}   \nonumber \\ 
   + &\ii N^{2}(k_{3}-k_{2})a_{3}\bar{a}_{2}\left[
\left(\frac{k_2}{\omega_2}-\frac{k_3}{\omega_3}\right)(k_3m_2-k_2m_3) \right]\mathfrak{P}_{3}\bar{\mathfrak{P}}_{2} \nonumber \\
   + &\ii N^{2}(k_{3}-k_{1})a_{3}\bar{a}_{1}\left[
\left(\frac{k_1}{\omega_1}-\frac{k_3}{\omega_3}\right)(k_3m_1-k_1m_3) \right]\mathfrak{P}_{3}\bar{\mathfrak{P}}_{1}+ \mathrm{c.c}., \label{eqn:NL_term2}
\end{align}
and  the third term is given by
\begin{align}
\frac{\partial \left(\{v,\psi\}\right)}{\partial z} = &\ii(m_1+m_2)f a_{1}a_{2}\left[
\left(\frac{m_1}{\omega_1}-\frac{m_2}{\omega_2}\right)(m_1k_2-m_2k_1) \right]\mathfrak{P}_{1}\mathfrak{P}_{2}   \nonumber \\    + &\ii(m_3-m_2)f a_{3}\bar{a}_{2}\left[
\left(\frac{m_3}{\omega_3}-\frac{m_2}{\omega_2}\right)(k_3m_2-k_2m_3) \right]\mathfrak{P}_{3}\bar{\mathfrak{P}}_{2}  \nonumber \\
   + &\ii(m_3-m_1)f a_{3}\bar{a}_{1}\left[
\left(\frac{m_3}{\omega_3}-\frac{m_1}{\omega_1}\right)(k_3m_1-k_1m_3) \right]\mathfrak{P}_{3}\bar{\mathfrak{P}}_{1}+ \mathrm{c.c}. \label{eqn:NL_term3}
\end{align}

\noindent In all expressions,  overbar denotes complex conjugate. There are additional terms with wavenumbers and frequencies different from that of the three waves initially assumed. These  are  non-resonant terms, which are  not important for resonant energy transfer, and hence are neglected.
\subsection{Amplitude evolution equations of a resonant triad}

From the  resonant terms at $\mathcal{O}(\epsilon^2)$ in \eqref{eqn:NL_term1}--\eqref{eqn:NL_term3}, we match those terms of the LHS and the RHS that have the same frequency and horizontal wavenumber. This finally leads to three amplitude evolution equations:
\begin{subequations}
\begin{align}
\frac{\partial a_{1}}{\partial t} + c_{x,1}^{(g)}\frac{\partial a_{1}}{\partial x} + c_{z,1}^{(g)}\frac{\partial a_{1}}{\partial z} + \mathcal{V}_1a_1 &= \frac{1}{2} \mathfrak{N}_{1}{a}_{3}\bar{a}_{2}{\exp(\ii \int_{-\infty}^{z}\Delta m\,dz  )}\label{eqn:wave1}\\
 \frac{\partial a_{2}}{\partial t} + c_{x,2}^{(g)}\frac{\partial a_{2}}{\partial x} + c_{z,2}^{(g)}\frac{\partial a_{2}}{\partial z} + \mathcal{V}_2a_2  &= \frac{1}{2} \mathfrak{N}_{2}{a}_{3}\bar{a}_{1}{\exp(\ii \int_{-\infty}^{z}\Delta m\,dz  )} \label{eqn:wave2}\\
  \frac{\partial a_{3}}{\partial t} + c_{x,3}^{(g)}\frac{\partial a_{3}}{\partial x} + c_{z,3}^{(g)}\frac{\partial a_{3}}{\partial z} + \mathcal{V}_3a_3  &= \frac{1}{2} \mathfrak{N}_{3}{a}_{1}{a}_{2}{\exp(\ii \int_{-\infty}^{z}-\Delta m\,dz  )}
\label{eqn:wave3}
\end{align}
\end{subequations}

 The functions $c_{x,j}^{(g)},c_{z,j}^{(g)}, \mathcal{V}_{j}$ and $\mathfrak{N}_{j}$ are given by:
 \begin{subequations}
 \begin{equation}
c_{x,j}^{(g)}(\epsilon_{n} z) \equiv \frac{k_{j}\left(N^{2}- \omega^{2}_{j}\right)}{\omega_{j}(k^{2}_{j}+m^{2}_{j})}, \hspace{0.2cm} c_{z,j}^{(g)}(\epsilon_{n} z) \equiv  -\frac{m_{j}(\omega_{j}^2-f^2)}{\omega_j(k_{j}^2+m_{j}^2)}, \hspace{0.2cm}  \mathcal{V}_j(\epsilon_{n} z) \equiv  \frac{\nu}{2}\left[k_j^2+m_j^2 + \frac{f^2m_j^2}{\omega^2_j} \right]
\label{eqn:group_speed}
\end{equation}
\begin{align}
\mathfrak{N}_{1}(\epsilon_{n} z) &\equiv  \frac{N^{2}(k_{3}-k_{2})}{k^{2}_{1}\omega_{1}+m^{2}_{1}\omega_{1}}\left[
\left(\frac{k_2}{\omega_2}-\frac{k_3}{\omega_3}\right)(k_3m_2-k_2m_3) \right]\left( \abs{ \frac{m_{1}}{{m_{2}m_{3}}} } \right)^{1/2} \nonumber \\ 
                  &-\frac{(\omega_{3}-\omega_{2})}{k^{2}_{1}\omega_{1}+m^{2}_{1}\omega_{1}}\left[(k_{3}m_{2}-k_{2}m_{3})\left(m^{2}_{3}+k^{2}_{3}-k^{2}_{2}-m^{2}_{2}\right) \right]\left( \abs{ \frac{m_{1}}{{m_{2}m_{3}}} }\right)^{1/2} \nonumber \\
             &-\frac{f^2(m_3-m_2)}{k^{2}_{1}\omega_{1}+m^{2}_{1}\omega_{1}} \left[\left(\frac{m_3}{\omega_3}-\frac{m_2}{\omega_2}\right)(k_3m_2-k_2m_3) \right]\left( \abs{ \frac{m_{1}}{{m_{2}m_{3}}} }\right)^{1/2}  ,\label{eqn:coupling_coefficient_1}
\end{align}
\begin{align}
\mathfrak{N}_{2}(\epsilon_{n} z) &\equiv  \frac{N^{2}(k_{3}-k_{1})}{k^{2}_{2}\omega_{2}+m^{2}_{2}\omega_{2}}\left[
\left(\frac{k_1}{\omega_1}-\frac{k_3}{\omega_3}\right)(k_3m_1-k_1m_3) \right]\left( \abs{ \frac{m_{2}}{{m_{1}m_{3}}}} \right)^{1/2} \nonumber \\ 
  &-\frac{(\omega_{3}-\omega_{1})}{k^{2}_{2}\omega_{2}+m^{2}_{2}\omega_{2}}\left[(k_{3}m_{1}-k_{1}m_{3})\left(m^{2}_{3}+k^{2}_{3}-k^{2}_{1}-m^{2}_{1}\right) \right]\left( \abs{ \frac{m_{2}}{{m_{1}m_{3}}} } \right)^{1/2} \nonumber \\
  &-\frac{f^2(m_3-m_1)}{k^{2}_{2}\omega_{2}+m^{2}_{2}\omega_{2}} \left[
\left(\frac{m_3}{\omega_3}-\frac{m_1}{\omega_1}\right)(k_3m_1-k_1m_3) \right]\left( \abs{ \frac{m_{2}}{{m_{1}m_{3}}} }\right)^{1/2} , 
  \label{eqn:coupling_coefficient_2}
\end{align}
\begin{align}
 \mathfrak{N}_{3}(\epsilon_{n} z) &\equiv  \frac{N^{2}(k_{1}+k_{2})}{k^{2}_{3}\omega_{3}+m^{2}_{3}\omega_{3}}\left[
\left(\frac{k_1}{\omega_1}-\frac{k_2}{\omega_2}\right)(k_1m_2-k_2m_1) \right]\left( \abs{ \frac{m_{3}}{{m_{2}m_{1}}} } \right)^{1/2} \nonumber \\
          &-\frac{(\omega_{1}+\omega_{2})}{k^{2}_{3}\omega_{3}+m^{2}_{3}\omega_{3}}\left[(k_{1}m_{2}-k_{2}m_{1})\left(m^{2}_{2}+k^{2}_{2}-k^{2}_{1}-m^{2}_{1}\right) \right]\left( \abs{ \frac{m_{3}}{{m_{2}m_{1}}} } \right)^{1/2} \nonumber \\
          &+ \frac{f^2(m_1+m_2)}{k^{2}_{3}\omega_{3}+m^{2}_{3}\omega_{3}} \left[\left(\frac{m_2}{\omega_2}-\frac{m_1}{\omega_1}\right)(k_2m_1-k_1m_2) \right] \left( \abs{ \frac{m_{3}}{{m_{2}m_{1}}} } \right)^{1/2}.
    \label{eqn:coupling_coefficient_3}
\end{align}
\end{subequations}
These equations generalize the ones obtained in \cite{lambnear} and \cite{bourget}  since in our case, the coefficients $c_{x,j}^{(g)},c_{z,j}^{(g)}, \mathfrak{N}_{j}$ and  $\mathcal{V}_{j}$ and are all dependent on the $z$-direction. 
The vector $(c_{x,j}^{(g)},\,c_{z,j}^{(g)})$ denotes the (weakly varying) group speed of the $j$-th wave, $\mathfrak{N}_j$ is the nonlinear coupling coefficient for the $j$-th wave, while 
$\mathcal{V}_{j}$ is the viscous term for the $j$-th wave. { The frequency of a wave is always considered positive, hence  the direction of  wave propagation is determined by the wave-vector ($k,m$). From the expression \eqref{eqn:group_speed}, we observe that positive (negative) $k_j$ implies propagation in the positive (negative) $x$ direction, and negative (positive) $m_j$ implies propagation in the positive (negative) $z$ direction.}

{We note here that \cite{grimshaw_1988,grimshaw_1994}  has derived the governing equations for a resonant triad in a more general setting of slowly varying   density stratification along with slowly varying background shear. Moreover, Grimshaw's theory also considers  slowly varying wavetrains for which both frequency and wavenumbers can have $\mathcal{O}(1)$ changes.} The primary focus in \cite{grimshaw_1988,grimshaw_1994}  was to analyse triad interactions near a critical layer (without any change in the background stratification), which is quite different from ours; we focus on the effects of slowly varying background stratification {on finite width `wave-packets' constituting a} triad. Moreover, we have included rotational effects and also derived the governing equations for amplitude evolution using a different method (WKB approximation), which allows explicit expressions for the phases.

Equations (\ref{eqn:wave1})--(\ref{eqn:wave3}) are the amplitude evolution equations of the waves, or `wave-packets' whose carrier waves satisfy the triad condition. {The length scales of the variation of amplitude function  and the variation of the  stratification function  are at least an order of magnitude higher than the length scale (vertical wavelength) of the waves.} {Notice from \eqref{eqn:eps_z} that the amplitude function and the stratification function may have the same length scales.}
For simplicity, \emph{we assume in all our subsequent studies the initial wave amplitudes to be independent of} $x$. {Moreover, we always assume that the width of the wavepacket in the $z$--direction is at least one order of magnitude greater than the wavelength of all three waves. {  Throughout the paper, the reduced order governing equations (\ref{eqn:wave1})--(\ref{eqn:wave3}) are applied to analyze settings for which the mismatch in the vertical wavenumber is always a small quantity (i.e. $\Delta m/m_{\textnormal{min}} \ll \mathcal{O}(1)$) over the entire physical space. Therefore the triads considered are global in nature, meaning, anywhere in the physical space the triads are always resonant or near-resonant. }} 

Since the evolution equations are themselves not capable of creating $x$ variations, amplitudes that are initially independent of $x$ remains so forever (i.e.\ evolves only along $z$). The functions  $c_{x,j}^{(g)},c_{z,j}^{(g)}, \mathfrak{N}_{j}$,  $\mathcal{V}_{j}$ and the exponential functions in the RHS of (\ref{eqn:wave1})--(\ref{eqn:wave3}) influence the energy transfer, and also create amplitude variations in the $z$--direction, even if the waves' amplitudes are initialized with no $z$--dependence. This is precisely due to the non-uniformity of the density stratification profile.
In fact, the origin of these exponential functions is the non-uniformity of the density stratification profile -- the vertical wavenumber does not satisfy the triad condition at all locations, which leads to  mismatch in the vertical wavenumber.
Thus, the argument of each exponential function represents the relative phase difference created between the waves (forming the triad) as they propagate through the non-uniformly stratified medium. Since such a mechanism introduces wave detuning (i.e. deviation from forming a resonant triad), 
hereafter we refer {the exponential function} as the \emph{detuning function}.

\subsection{Energy evaluation}

The evolution of energy for these three waves is calculated by considering the total energy (kinetic + potential), where total energy density at an instant is given by:
\begin{equation}
\widehat{\textnormal{TE}}_{j} \equiv \frac{\rho_{0}}{2}\left( u^{2}_{j} + v_{j}^2 + w^{2}_{j} \right) + \frac{\rho_{0}}{2}\left( \frac{b_j^2}{N^2} \right) = \frac{\rho_{0}}{2}\left[\left( \frac{\partial \psi_{j}}{\partial z}\right)^{2} + \left(\frac{\partial \psi_{j}}{\partial x}\right)^{2} + v_{j}^{2} + 
\frac{b^2_j}{N^2}\right].
\label{eqn:KE_def}
\end{equation}
\noindent The time averaged total energy density for an internal gravity wave over its time period is given by: 
\begin{equation}
\langle \widehat{\textnormal{TE}}_{j} \rangle \equiv \frac{\omega_{j}}{2\pi}\int_{0}^{{2\pi}/{\omega_{j}}} \frac{\rho_{0}}{2}\left[\left( \frac{\partial \psi_{j}}{\partial z}\right)^{2} + \left(\frac{\partial \psi_{j}}{\partial x}\right)^{2} + v_{j}^{2} +
\frac{b^2_j}{N^2} \right] dt.
\label{eqn:KE_int_time}
\end{equation}
The total energy in the domain is calculated by integrating in the $z$-direction:
\begin{equation}
\textnormal{TE}_{j} \equiv \int_{0}^{H}\langle \widehat{\textnormal{TE}}_{j}\rangle dz = \intop_{0}^{H} 2\rho_{0}\left[\frac{\omega^{2}_jk^{2}_{j} + (f^2+\omega^{2}_j)m^{2}_{j} + k^2_jN^2}{\omega_j^2 m_{j}}\right] a_{j}\bar{a}_{j}  dz,
\label{eqn:KE_averaged}
\end{equation}
where $H$ is the length of the domain in the $z$-direction. We non-dimensionalize $\textnormal{TE}_{j}$ with the initial energy  of wave `3':  
 \begin{equation}
 {E}_{j} \equiv \frac{\textnormal{TE}_{j}}{\,\,\,\,\,\,\textnormal{TE}_{3}|_{t = 0}}.
\label{eqn:KE_nondimesion}
 \end{equation}



\section{Normal mode analysis for interaction of inviscid detuned plane waves in uniform stratification
\label{Section:3}}
Energy transfer between finite width internal gravity wave beams is an important problem in oceanography; previous studies  \citep{bourget_width_2014,karimi_2014} have shown that the width of the primary internal gravity wave plays a key role in the energy transfer process. The daughter waves should spatially overlap with the parent wave for a given amount of time so that they can exchange energy effectively.
The overlap time between  different beams is primarily dependent on the group speed (apart from the individual beam width) of the internal wave beams. However, even for plane  waves (i.e. packets of infinite width) or wave-packets having large width which do not move out of each other's range,  group speed can play a key role in deciding the growth rates of the daughter waves, provided there is a spatial variation in  amplitude profile of any of the constituent waves \citep{pumpcraik}.
{In this section, we use normal mode analysis to estimate the growth rates of the daughter waves in uniform stratification without viscosity, where all three waves can have different vertical group speeds.}
To this end, let us consider the inviscid governing equations for a triad with a constant wavenumber mismatch: 
\begin{subequations}
\begin{align}
\frac{\partial a_{1}}{\partial t} + c_{z,1}^{(g)}\frac{\partial a_{1}}{\partial z} & = \frac{1}{2} \mathfrak{N}_{1}{a}_{3}\bar{a}_{2}\ee^{\ii \Delta m z} \label{eqn:GDE_constant_mismatch_1},\\
\frac{\partial a_{2}}{\partial t} + c_{z,2}^{(g)}\frac{\partial a_{2}}{\partial z} &  = \frac{1}{2} \mathfrak{N}_{2}{a}_{3}\bar{a}_{1}\ee^{\ii \Delta m z} \label{eqn:GDE_constant_mismatch_2},\\
\hspace{0.5cm}\frac{\partial a_{3}}{\partial t} + c_{z,3}^{(g)}\frac{\partial a_{3}}{\partial z}  & = \frac{1}{2} \mathfrak{N}_{3}{a}_{1}{a}_{2}\ee^{- \ii \Delta m z}. \label{eqn:GDE_constant_mismatch_3}
\end{align}
\end{subequations}
Here $\Delta m \equiv m_{3}-m_{1}-m_{2}$ is the mismatch in the vertical wavenumber (which introduces the \emph{detuning}), furthermore  {$\Delta m/m_{j} \sim \mathcal{O}(\epsilon_z)$} is assumed.
In PSI, usually the parent wave's amplitude (here it is wave `3') is very large in comparison to the two daughter waves. Hence the nonlinear term in (\ref{eqn:GDE_constant_mismatch_3}) is negligible in the initial stages of the problem {(i.e., the equation follows the scaling of \eqref{eqn:scaling_analysis_2})}. This is known as the \emph{pump-wave approximation} \citep{pumpcraik}. 
Thus, under the pump wave approximation, (\ref{eqn:GDE_constant_mismatch_1})--(\ref{eqn:GDE_constant_mismatch_3}) reduces to coupled linear PDE.  


Under the pump wave approximation, an oscillatory solution for $a_3$ in (\ref{eqn:GDE_constant_mismatch_3}) is possible. Hence we assume that: $a_{3} = A_{3} \ee^{\ii  M_{3} ( z-c_{z,3}^{(g)}t)}$, and similarly the solution for $a_{1}$ and $a_{2}$ is assumed as: $a_{1} = \Tilde{a}_{1}(\epsilon_{t} t)\ee^{\ii (M_{1}z - M_{3}c_{z,3}^{(g)}t)}$ and $a_{2} = \Tilde{a}_{2}(\epsilon_{t} t)\ee^{\ii M_{2}z}$. Here $M_{j}$ vertical wavenumber for the amplitude profile (not to be confused with vertical wavenumbers, $m_j$).  $A_{3}$ is a constant denoting the amplitude of the  wave `3' (the pump wave).

The relation between $M_{1}, M_{2}, M_{3},$ and $\Delta m$ is then given by:  $M_{2} = M_{3} + \Delta m -M_{1}$, {which makes the problem variable separable}. Substituting these  in (\ref{eqn:GDE_constant_mismatch_1})--(\ref{eqn:GDE_constant_mismatch_3}) reduces the governing equations to:
\begin{subequations}
\begin{align}
\frac{\partial \Tilde{a}_{1}}{\partial t} + \ii (M_{1} c_{z,1}^{(g)}-M_{3}c_{z,3}^{(g)}) \Tilde{a}_{1} & = \frac{1}{2} \mathfrak{N}_{1}A_{3}\bar{\Tilde{a}}_{2}, \label{eqn:GDE_constant_mismatch_pump_1_time} \\
\frac{\partial \Tilde{a}_{2}}{\partial t} + \ii (M_{3}+\Delta m-M_{1})c_{z,2}^{(g)} \Tilde{a}_{2} &  = \frac{1}{2} \mathfrak{N}_{2}A_{3}\bar{\Tilde{a}}_{1}. \label{eqn:GDE_constant_mismatch_pump_2_time}
\end{align}
\end{subequations}
If we consider the solution of $\Tilde{a}_{j}$ (where $j=1,2$) to be of the form: $\Tilde{a}_{j} = \ee^{-\ii\Omega_j t}$, then the growth rate of the $\Tilde{a}_{j}$ is defined as: $\textnormal{GR}_{j} \equiv  \Im(\Omega_j)$. The amplitude growth rates are then found to be:
 \begin{equation}
 \textnormal{GR}_1 = \textnormal{GR}_2=\frac{1}{2}\sqrt{\gamma_{A}-\gamma_{M}}.
\label{eqn:Growth_rates_general}
 \end{equation}
where 
\begin{equation}
\gamma_{A}\equiv \mathfrak{N}_{1}\mathfrak{N}_{2}A_{3}^{2} \hspace{0.5cm} \textnormal{and} \hspace{0.5cm} \gamma_{M}\equiv\left\{M_{1} c_{z,1}^{(g)}-M_{3}c_{z,3}^{(g)}+(M_{3}+\Delta m-M_{1})c_{z,2}^{(g)}\right\}^{2}.
\label{eqn:gammaa_and_gammam}
\end{equation}
For maximum growth rates, we must have $\gamma_{M} = 0.$
\begin{equation}
 \gamma_{M} = 0.
 \label{eqn:condition_max_growth_general}
\end{equation}
Here $\gamma_A$ represents the nonlinear forcing due to the parent wave and $\gamma_M$ represents the growth reduction due to the spatial variation of the waves in triad. 
{We emphasize here that the above condition for obtaining maximum growth rates is quite general since it allows all three waves in the triad to have a spatial variation as well as a wavenumber mismatch.} We also note here that special cases of the condition that we derived have been explored previously. For example, \cite{pumpcraik} studied the parameter space where $c_{z,1}^{(g)} = c_{z,2}^{(g)}$ when $M_{3} = 0$, in which case a detuned triad cannot have the same growth rate as a resonant triad; as the detuning is increased, the growth rate keeps on decreasing for any normal mode form of $a_{1}$ and $ a_{2}$. {In addition, \cite{mcewan} explored the parameter space when $c_{z,1}^{(g)}\neq c_{z,2}^{(g)}$ and $M_{3}=0$.}\\

\begin{figure}
{\centering
\includegraphics[width=1.0\linewidth,keepaspectratio]{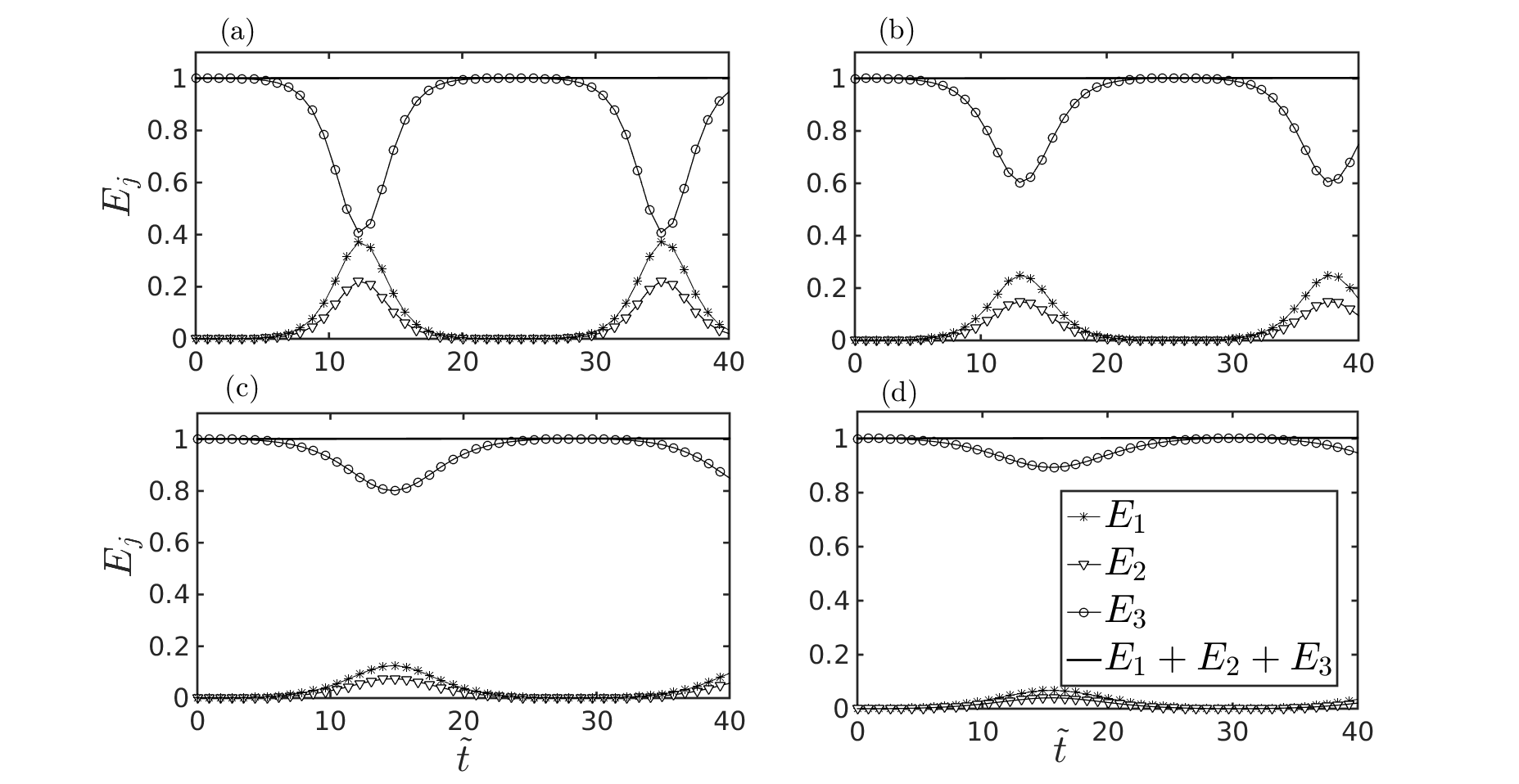}}
\caption[Energy evolution plot for triad case 4]{{ Evolution of non-dimensional energy of each wave with time. (a) $\gamma_{M}/\gamma_{A} = 0.4$, (b) $\gamma_{M}/\gamma_{A} = 0.6$, (c) $\gamma_{M}/\gamma_{A} = 0.8$, and (d) $\gamma_{M}/\gamma_{A} = 0.9$.  The non-dimensional time, $\tilde{t}$ is defined as $\tilde{t} \equiv t\sqrt{\gamma_{A}}$.}}
\label{fig:normal_mode}
\end{figure}
Even though certain normal modes have exponential growth, the primary  wave still need not transfer its energy completely to the daughter waves. This is dependent on the parameter $\gamma_{M}/\gamma_{A}$. From  (\ref{eqn:Growth_rates_general}), it can be seen that when $\gamma_{M}/\gamma_{A} > 1$, the particular normal mode would not grow exponentially, exponential growth would occur only when $\gamma_{M}/\gamma_{A} < 1$. We perform numerical experiments where we vary the parameter $\gamma_{M}/\gamma_{A}$ to see its importance in energy transfer in between the waves. 
A triad is used  with daughter waves (wave-1 and wave-2) and parent wave (wave-3) having initial amplitudes such that $|a_1|/|a_3| = 0.012$ and $|a_2|/|a_3| = 0.015$ for all the simulations. The following frequencies and wavenumbers are chosen:  $\omega_{1} = 0.10N$, $\omega_{2} = 0.18N$, $\omega_{3} = 0.28N$,  and $k_{1}H = 0.31$, $k_{2}H = -0.9$, $k_{3}H = -0.59$. Here $N = 10^{-3} \textnormal{s}^{-1}$ and $H = 100 \textnormal{m}$ is used. The results are shown in figure \ref{fig:normal_mode}. \\
In these {simulations}, even though the normal modes have exponential growth the parent wave does not completely exchange its energy. In figure \ref{fig:normal_mode}(d), when $\gamma_{M}/\gamma_{A} = 0.9$, the parent wave transfers only $\approx 10\%$ of its total energy, while for $\gamma_{M}/\gamma_{A} = 0.8$, it is $\approx 20\%$ which is shown in figure \ref{fig:normal_mode}(c). A pattern can be noticed here -- the maximum percentage of energy lost by the parent wave can be given by $100\times(1-\gamma_{M}/\gamma_{A})$ {(same holds for figures \ref{fig:normal_mode}(a) and \ref{fig:normal_mode}(b))}. Hence using normal mode analysis in uniform stratification, the  maximum amount of energy the parent wave exchanges with the daughter waves can also be predicted. Even though a specific triad is used here, similar behavior is also observed for other triads.

In summary, even when the daughter waves undergo exponential growth (using normal mode assumption), complete energy transfer to the daughter waves is not possible when $\gamma_{M} \neq 0$. The maximum energy transferred from the primary to the daughter waves in the later stages can be accurately predicted by $\gamma_M/\gamma_A$, which, in fact, can be estimated from the initial conditions.

\section{Interaction between  wave-packets in uniform stratification under resonant and detuned conditions \label{Section:4}}

 In this section, we focus on the energy transfer between resonant as well as near-resonant inviscid wave-packets in uniform density stratification. In \S \ref{Section:3} we already showed that in general, as the detuning of plane waves forming a triad is increased, the growth rate of the daughter waves get decreased; see \eqref{eqn:Growth_rates_general}. We study the effect of detuning (or mismatch) in vertical wavenumber condition on the energy transfer between wave-packets. Throughout this section, the parent wave is considered as a wave-packet of finite width. \cite{mcewan} explored the parameter space where the parent wave-packet was of infinite width (plane wave) while the daughter waves were a finite sized wave-packet. Since in our case, the parent wave also has a finite width (hence a parent wave-packet), the group speed of the parent wave-packet becomes important. The governing equations considered in this section are  \eqref{eqn:GDE_constant_mismatch_1}--\eqref{eqn:GDE_constant_mismatch_3}. 

The three evolution equations  
are solved using Runge Kutta 4 method in time and second order accurate discretization scheme for the term ${\partial a_j}/{\partial z}$, where the scheme is forward or backward depending on the group speed direction of the particular wave.
Throughout this section, the initial amplitude profile for all the three waves forming the triad is chosen to be Gaussian shape in the $z$-direction:
\renewcommand{\theequation}{\arabic{section}.\arabic{equation}a,b,c}
\begin{equation}
a_{1} = A_{1} \ee^{-\left( z/W_{p(1)}\right)^{2}}, \hspace{1cm} a_{2} = A_{2} \ee^{-\left( z/W_{p(2)}\right)^{2}}, \hspace{1cm} a_{3} = A_{3} \ee^{-\left( z/W_{p(3)} \right)^{2}}.
\label{eqn:amp_def_gauss_1}
\end{equation}
\renewcommand{\theequation}{\arabic{section}.\arabic{equation}}
Before solving equations \eqref{eqn:GDE_constant_mismatch_1}--\eqref{eqn:GDE_constant_mismatch_3}  numerically, we define two non-dimensional numbers $\Pi_w$ and $\Pi_{m}$, which will be shown to play crucial role in the energy transfer process:
\begin{equation}
    \Pi_{w} \equiv \abs{ \frac{W_{p(3)}\sqrt{\mathfrak{N}_{1} \mathfrak{N}_{2} A_3^2}}{c_{z,3}^{(g)}}}, \hspace{1cm} \Pi_{m} \equiv \abs{\frac{\sqrt{\mathfrak{N}_{1} \mathfrak{N}_{2} A_3^2}}{\Delta m \, c_{z,3}^{(g)}}}.
    \label{eqn:non_d_s4}
\end{equation}

These two non-dimensional numbers, $\Pi_m$ and $\Pi_w$, are very similar to $\gamma_M/\gamma_A$ defined in \S \ref{Section:3}. In $\Pi_m$, the length scale is decided by the detuning ($\Delta m$). In $\Pi_w$, the length scale is decided by the width of the wave-packets. Systems with $\Pi_m \rightarrow \infty$ imply interaction between waves with no detuning ($\Delta m =0$). In systems where $\Pi_w \gg \mathcal{O}(1)$, the wave-packets have enough time to interact and exchange energy. On the contrary, systems where $\Pi_w \ll \mathcal{O}(1)$ imply wave-packets moving out of each others' range before they can exchange energy.  Increasing $\Pi_w$ by increasing the width of the packets will not result in an increase in the growth rate of the daughter wave-packets beyond a maximum value given by ${\sqrt{\mathfrak{N}_{1} \mathfrak{N}_{2} A_3^2}}/2$ (i.e. the growth rate of plane wave triads, which can be considered as wave-packets of infinite width). 
The difference between near-resonant and resonant wave-packet interaction is negligible when $\Pi_m \gg \mathcal{O}(1)$ for any value of $\Pi_w$.  However, the difference between resonant and near-resonant wave-packet interaction is significant for  $\Pi_m \ll \mathcal{O}(1)$ for $\Pi_w \sim \mathcal{O}(1)$. {This is shown by the numerical experiments below.}

For studying the energy transfer between detuned wave-packets forming a triad, we fix their group speeds and nonlinear coefficients, however the width of the wave-packets are varied. Moreover, for each wave-packet width, the detuning between the waves is slowly varied and the effect of this detuning on the energy transferred to the daughter wave-packets is studied. We emphasize here that in realistic systems,  variation in background stratification is needed to cause a detuning of vertical wavenumber. This would lead to varying nonlinear coefficients and group speeds, which would in turn make it difficult to underpin the key role played by detuning alone. To circumvent this issue, we keep background stratification as constant (hence group speeds and nonlinear coefficients are constant), but independently vary the detuning.
To this end, the following frequencies and wavenumbers are chosen:  $\omega_{1} = 0.10N$, $\omega_{2} = 0.18N$, $\omega_{3} = 0.28N$,   $k_{1}H = 0.31$, $k_{2}H = -0.9$, $k_{3}H = -0.59$, where $N = 10^{-3} \textnormal{s}^{-1}$ and $H = 100 \textnormal{m}$. 
We define the amplitudes following (\ref{eqn:amp_def_gauss_1}), with  $A_{1} = A_{2} = 10^{-5}$ $\textnormal{m}^{5/2}\textnormal{s}^{-1}$ and $A_{3} = 10^{-2}$ $\textnormal{m}^{5/2}\textnormal{s}^{-1}$ (wave-3's energy is much more than the other two waves). In all simulations, $W_{p(1)} = W_{p(2)} = W_{p(3)}$ is assumed. This resulting triad system is similar to that of PSI. 
The quantity $\Delta m$ is non-dimensionalized with the parent wave's vertical wavenumber ($m_{3}$), and  $\Delta m/m_{3}$ is varied between $0$ and $0.1$ for all the different wave-packet sizes used. The wave-packet sizes chosen for this analysis are $W_{p(1)} = 30\lambda_3, 60\lambda_{3}, 120\lambda_{3}$ and $ 240\lambda_{3}$. For the wave-packet size $W_{p(1)} =  240\lambda_{3}$, we have $\Pi_w = 50$. Furthermore, for a detuning of $\Delta m/m_3 = 0.1$, we have $\Pi_m=0.34$.

\begin{figure}
{\centering
\includegraphics[width=1.0\linewidth,keepaspectratio]{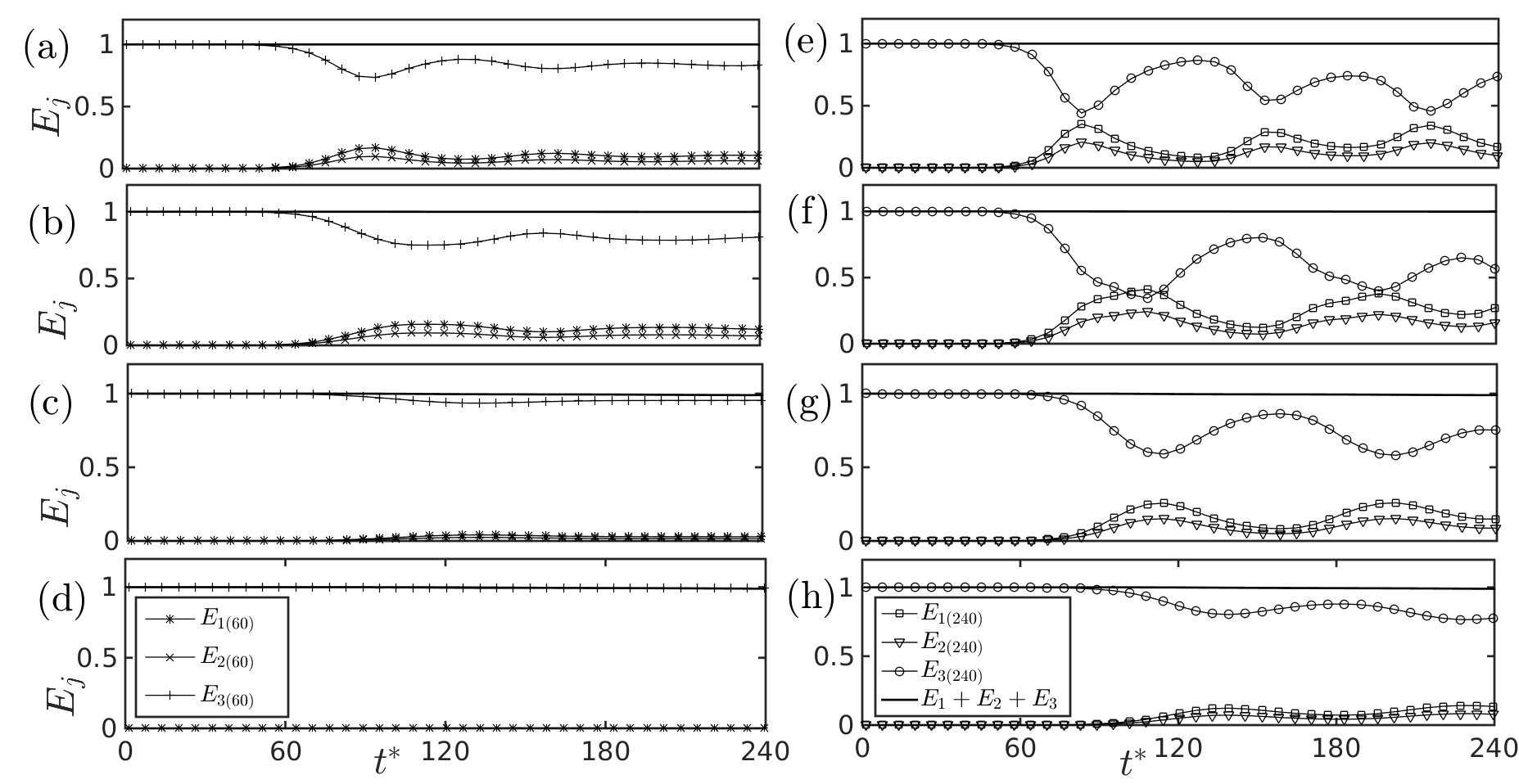}}\caption[Energy evolution plot for triad case 4]{{ Energy evolution plots for $\Delta m/m_{3} = 0.00$ (a,e) (resonant),  $\Delta m/m_{3} = 0.04$ (b,f), $\Delta m/m_{3} = 0.08 $ (c,g), $\Delta m/m_{3} = 0.1$ (d,h). The abscissa $t^{*}=t\omega_{3}/2\pi$ represents non-dimensional time. Two different wave-packet sizes are considered: (i) $W_{p(1)}=W_{p(2)}=W_{p(3)}=60 \lambda_{3}$, for (a), (b), (c), and (d) and (ii) $W_{p(1)}=W_{p(2)}=W_{p(3)}=240 \lambda_{3}$, for (e), (f), (g) and (h).}}
\label{fig:finitepacketdetuningeffect}
\end{figure}

The effect of detuning on the energy transfer among the wave-packets is shown in figure \ref{fig:finitepacketdetuningeffect} for two different wave-packet sizes: (i) $W_{p(j)} = 60\lambda_3$ and (ii) $W_{p(j)} = 240\lambda_3$, where $j=1,2,3$.  When $W_{p(1)}=60\lambda_{3}$, the parent wave-packet in the resonant case transferred $30\%$ of its total energy, while the transfer was less than $1\%$ for $\Delta m/m_{3}=0.1$. Hence  detuning  may act as an extra constraint in the energy transfer  between wave-packets. 
An interesting fact occurs for the wave-packet size of $W_{p(1)} = 240\lambda_{3}$ -- the energy exchange corresponding to $\Delta m/m_{3} = 0.04$ is more than the resonant wave-packet at a certain point of time; compare figure \ref{fig:finitepacketdetuningeffect}(f) with  figure \ref{fig:finitepacketdetuningeffect}(e). Putting quantitatively, the parent wave for the resonant case  
transferred $56\%$ of its total energy at $t^{*}=84$, however it  transferred $66\%$ of its total energy at $t^{*}=106$ when $\Delta m/m_{3}=0.04$. This is because in the case of no detuning (i.e. resonant condition), the wave-packets exchange energy faster than the detuned packets. The energy transfer near the peak region of the Gaussian bump (in comparison to the flank regions) of the parent wave-packet's amplitude profile is so fast that at $t^{*}\approx 84$, the direction of energy transfer in that particular region reverses, that is, the parent wave starts gaining energy near the `peak' of the Gaussian region. Meanwhile the flank regions of the parent wave-packet still provides energy to the daughter waves.
Hence the net energy exchange of the daughter wave-packets become near zero (near $t^{*}=80$); see figure \ref{fig:finitepacketdetuningeffect}(e).

As $t^{*}$ further increases, the daughter waves provide more energy to the parent waves than it takes away, therefore the net energy of the parent wave-packet increases. The time for reversal of energy transfer (daughter wave-packets providing energy to the parent wave-packet) is smaller for a resonant case than the detuned cases. Meanwhile for a detuned case, the reversal of energy transfer near the top region of the Gaussian bump {(of the amplitude profile of the parent wave-packet)} is slower, which results in outer regions of the Gaussian bump transferring more energy (before the reversal of energy transfer) in comparison to the resonant packet.
To see this in more detail, at $t^{*} = 106$ of figure \ref{fig:finitepacketdetuningeffect}(e), the parent wave has transferred around $80\%$ of its energy to the daughter waves ($E_{3} = 0.2$), if we exclude the energy which is returned back from the daughter waves. At the same $t^{*}$ for $\Delta m/m_{3} = 0.04$ (figure \ref{fig:finitepacketdetuningeffect}(f)),
the parent wave has transferred around $71\%$ of its energy to the daughter waves $(E_{3} = 0.29)$ excluding the energy transferred back from the daughter waves. Hence the key reason behind a parent wave-packet under detuning condition transferring more energy under resonant condition is due to the fact that in the latter case, a reversal of energy transfer occurs near the peak of the Gaussian bump in the parent wave's amplitude profile. For $W_{p(1)} = 30\lambda_{3}$, the parent wave-packet exchanged ({for all the values of $\Delta m$}) only about $1\%$ of its total energy at best. For wave-packet size of $W_{p(1)} = 120 \lambda_{3}$, for all values of $\Delta m$, the parent wave-packet  transferred more (less) energy than $W_{p(1)} = 60\lambda_3$ ($W_{p(1)} = 240\lambda_3$). 
 
\begin{figure}
{\centering
\includegraphics[width=1.0\linewidth,keepaspectratio]{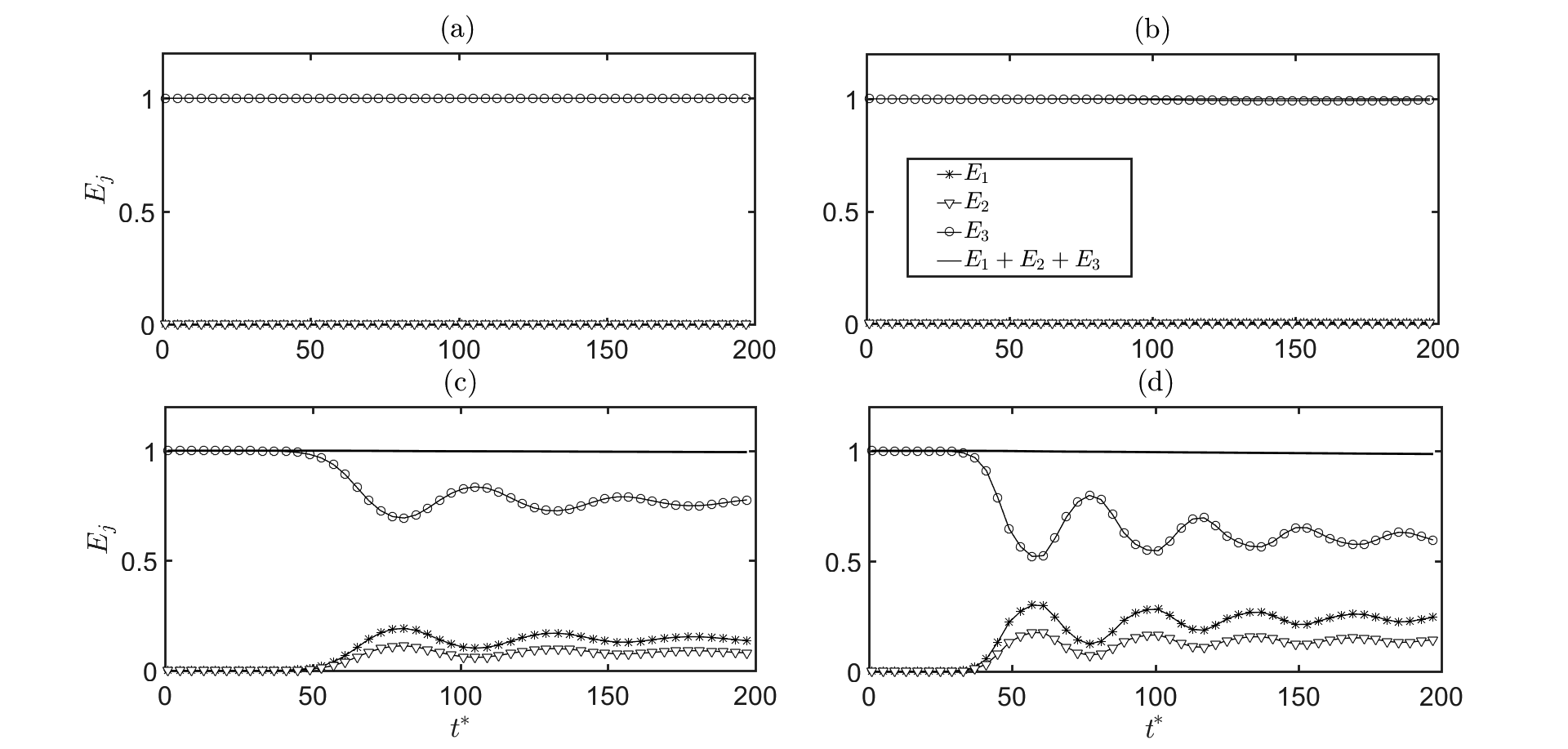}}\caption[Energy evolution plot for triad case 4]{{ Energy evolution plots for $\Delta m/m_{3} = 0.1$ and $W_{p(1)} = 60\lambda_3$. The parent wave amplitude used in the sub-figures: (a) $A_3 = 0.005$ (b) $A_3 = 0.01$, (c) $A_3 = 0.015$, (d)$A_3 = 0.02$.}}
\label{fig:finitepacketdetuningeffect_amp}
\end{figure}

For $\Delta m/m_{3} = 0.1$ and $A_3=0.01$, increasing the packet size beyond $W_{p(1)} = 240\lambda_{3}$ did not result in increased rate of energy transfer to daughter wave-packets. For example, the case of $W_{p(1)} = 960\lambda_{3}$ with $\Delta m/m_{3} = 0.1$ lost approximately $20\%$ of its total energy at $t^{*} \approx 130$ (similar to the case of $W_{p(1)} = 240\lambda_{3}$). Moreover, in the case of $W_{p(1)} = 240\lambda_{3}$, the parent wave-packet exchanged around $40\%$ of its total energy around $t^{*} = 350$ {(for $\Delta m =0$ case and same sized wave-packets, the parent wave-packet exchanged $\approx 56 \%$ of its energy at $t^{*}=82$)}. These parameters fall in the regime $\Pi_w \gg \mathcal{O}(1)$ and $\Pi_w \ll \mathcal{O}(1)$ -- even though the wave-packets have enough time to interact, the detuning reduces the rate of energy transfer between the packets in comparison to the resonant  case. 

 We study the variation in energy transfer with parent wave's amplitude ($A_{3}$), keeping the detuning fixed at $\Delta m/m_{3} = 0.1$, and the wave-packet widths at $W_{p(1)}=W_{p(2)} =W_{p(3)}=60\lambda_3$. All other parameters are also kept constant. 
Increasing $A_{3}$ increases the percentage of energy transferred from the parent wave-packet to the daughter wave-packets; see figure \ref{fig:finitepacketdetuningeffect_amp}. More importantly, increasing $A_{3}$ also increases the rate of energy transfer. This behavior is  consistent for all values of $\Delta m/m_3$. Hence the effect of detuning is continuously reduced as the parent wave's amplitude is increased. Furthermore, for wave-packets with $\Pi_m \gg \mathcal{O}(1)$, beyond a certain amplitude of the parent wave,  detuning has negligible effect on the energy transfer.

To summarize, energy transfer (from primary to daughter) in finite-width wave-packets
is monotonically affected as the detuning increases. The  width of the wave-packets have to be larger for the detuned case than the resonant case in order to exchange the same percentage of energy, when $\Pi_m \sim \mathcal{O}(1)$.

\section{Interactions between wave-packets in weakly varying stratifications }\label{Section:5}

\subsection{Interacting inviscid wave-packets in weakly varying stratification}\label{section:5.1} 


In this subsection we focus on wave-packets exchanging energy in weakly varying stratification. Energy transfer of finite width wave-packets in weakly non-uniform stratification, without considering viscosity, is mainly affected by four factors:
\begin{enumerate}
    \item  Change in the width (or length scale) of the wave-packets.
    \item  Varying vertical group speed ($c_{z,j}^{(g)}$) of the wave-packets (as shown in \S \ref{Section:4}, group speed  is key in deciding the effect of detuning between the waves).
    \item  Detuning ($\Delta m$).
    \item  Nonlinear coupling coefficients ($\mathfrak{N}_{j}$).
\end{enumerate}

\subsubsection{Effect on vertical group speed and wave-packet size when packets move to a different stratification} \label{section:5.1.1}
When a wave-packet travels from one background stratification to a different stratification, its vertical group speed changes.  
Furthermore, the width of the wave-packet also changes. 
The angular frequency and horizontal wavenumber, however, remains unchanged.  
The inviscid governing equation for the amplitude of a wave-packet moving through a non-uniform stratification is given by:
\begin{equation}
    \frac{\partial a_j}{\partial t} + c_{z,j}^{(g)}(\epsilon_n z)\frac{\partial a_j}{\partial z} = 0,
    \label{eqn:packet_movement}
\end{equation}
where $c_{z,j}^{(g)}(\epsilon_n z) \equiv -{m_{j}(\omega_{j}^2-f^2)}/{\omega_j(k_{j}^2+m_{j}^2)}$ is the vertical direction group speed of the packet, which is a function of stratification. Here we always assume that the wave-packet's energy is completely transmitted across the variable stratification. This is a reasonable assumption when the length scale of stratification's variation with space is much larger than the wave's vertical wavelength (\cite{mathur_2009}). Hence the energy of the wave-packet (given by \ref{eqn:KE_averaged}) will be constant as it moves through the varying stratification. To study how the wave-packets' size varies, we assume any arbitrary function for the amplitude ($a_j$) at $t=0$, which is given by:
\begin{equation}
a_j(z,0) = F(z),
\label{eqn:amp_ini}
\end{equation}
Let us assume this particular wave-packet travels from a constant stratified region, where the group speed is $c_{z,j}^{(g,1)}$, to another constant stratification region, where the group speed is $c_{z,j}^{(g,2)}$ (and the  {rate of} stratification variation is slow).  
The wave-packet's shape at any time $t$ in this new region is simply given by:
\begin{equation}
a(z,t) = F(\hat{z}) \hspace{1cm} \textnormal{where} \hspace{1cm} \hat{z} \equiv z{c_{z,j}^{(g,1)}}/{c_{z,j}^{(g,2)}}.
\label{eqn:amp_fin}
\end{equation}
The length scale of the packet has been re-scaled corresponding to the ratio of the group speed in the two regions. Thus, using the definition for group speed (given by \eqref{eqn:group_speed}), it can be straightforwardly concluded that wave-packet of a given size moving from a lower (higher) to a higher (lower) stratification will have its width reduced (increased). An important point worth noticing is that, even though the size of the wave-packet decreases (increases) in higher (lower) stratification, its group speed also decreases (increases) by the same factor as shown in (\ref{eqn:amp_fin}).  Hence the interaction time-scale among wave-packets would remain unchanged with the change in stratification. For any $\omega/N$ ratio, the group speed of a wave-packet always decreases (increases) when the packet moves to a higher (lower) stratification. For waves having  $\omega\ll N$, the group speed is inversely proportional to the background stratification, as shown below:
\begin{equation}
  c_{z,j}^{(g)} =  -\frac{m_{j}(\omega_{j}^2-f^2)}{\omega_j(k_{j}^2+m_{j}^2)} = -\frac{(\omega_{j}^2-f^2)^{3/2}}{\omega_j k_{j}\sqrt{(N^{2}-\omega_j^2)}} \approx -\frac{(\omega_{j}^2-f^2)^{3/2}}{\omega_j k_{j}N}. 
\label{eqn:GS_explain}
\end{equation}

\subsubsection{Mismatch in the vertical wavenumber condition when wave-packets move to a different stratification} \label{section:5.1.2}

\begin{figure}
{\centering
\includegraphics[width=1.0\linewidth,keepaspectratio]{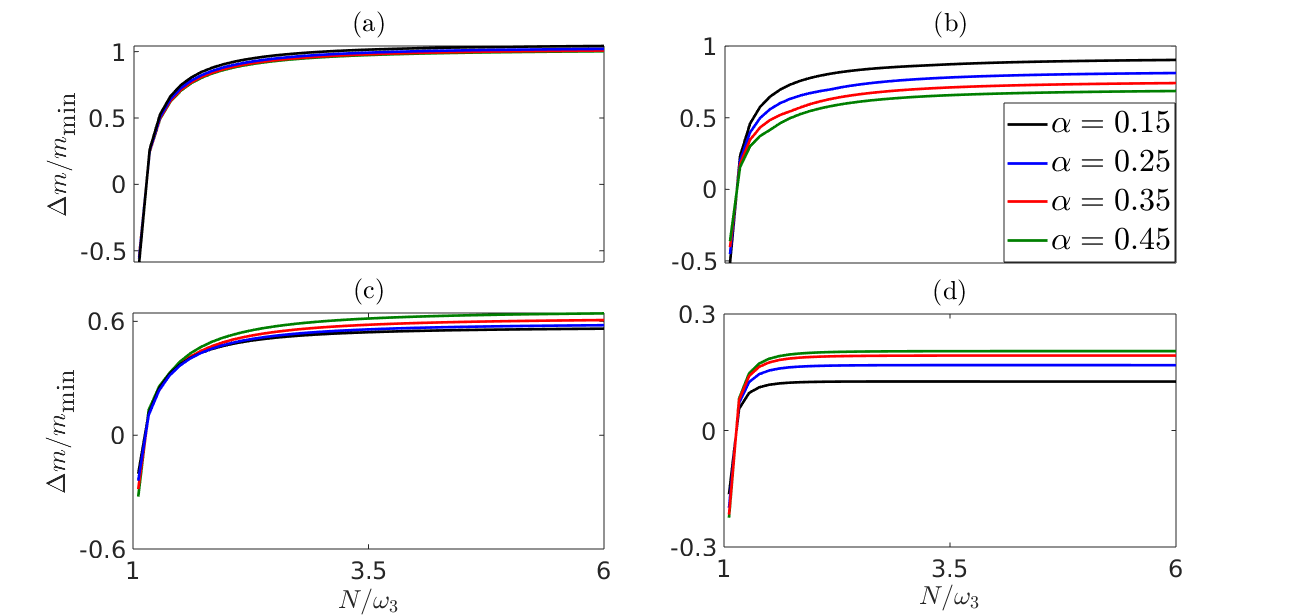}}
\caption[Energy evolution plot for triad case 4]{ Variation of detuning  for $\omega_3/N_b = 0.9$ and various daughter wave combinations. (a) {$(k_1/k_3,|{m_1/m_2}|)\in (1,\infty)\times (0,1)$}, (b) {$(k_1/k_3,|{m_1/m_2}|)\in (0,1)\times (0,1)$}, (c) {$(k_1/k_3,|{m_1/m_2}|)\in (0,1)\times (1,\infty)$}, and (d) {$(k_1/k_3,|{m_1/m_2}|)\in (1,\infty)\times (1,\infty)$}. Rotational effects are neglected.}
\label{fig:mismatchnvariation_differnt_triad}
\end{figure}

{When a wave-packet travels from one stratification to another, its vertical wavenumber changes, as evident from (\ref{eqn:vertical_wavenumber}). Therefore, if three wave-packets form a resonant triad on a particular background stratification, they will fail to do so once they move to another region with a different background stratification -- there will be a detuning ($\Delta m$) of the vertical wavenumbers. The main factors which influence detuning are the waves' frequencies, wavenumbers and the background stratification, whose effect is elaborated in  figures \ref{fig:mismatchnvariation} and \ref{fig:mismatchnvariation_differnt_triad}. 

 \begin{figure}
{\centering
\includegraphics[width=1.0\linewidth,keepaspectratio]{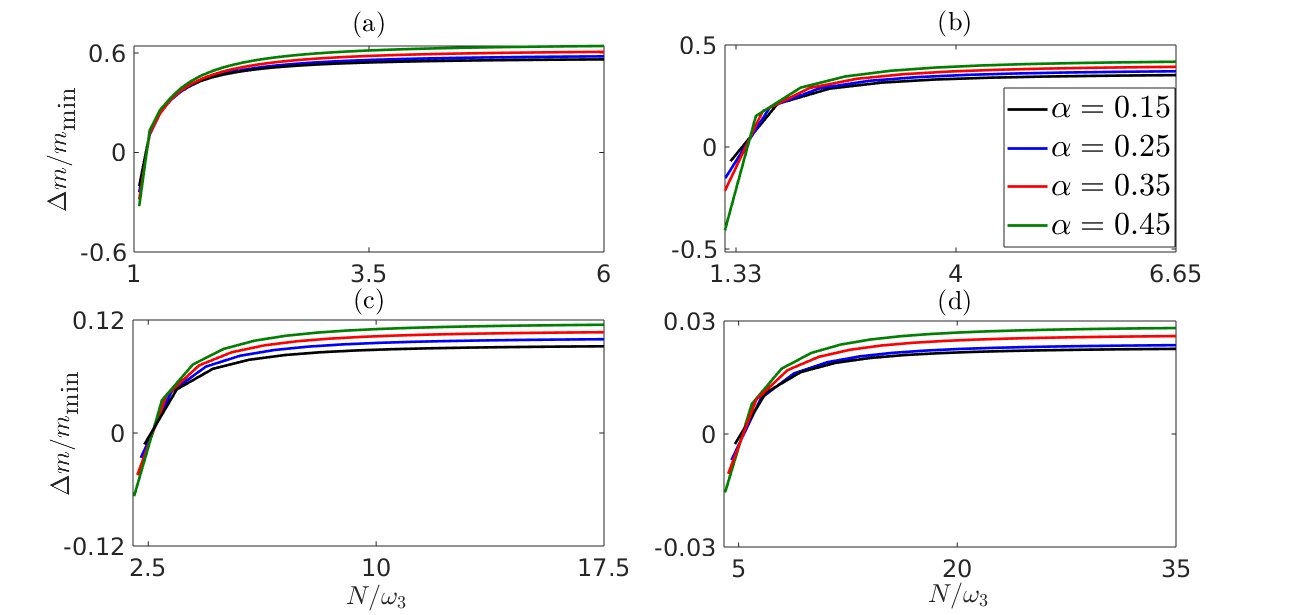}} 
\caption[Energy evolution plot for triad case 4]{ Variation of detuning in vertical wavenumber for different values of $\omega_3/N_b$. (a) $\omega_{3}/N_b = 0.9$, (b) $\omega_{3}/N_b = 0.75$, (c) $\omega_{3}/N_b = 0.4$ and (d) $\omega_{3}/N_b = 0.2$. Here $m_{\textnormal{min}}$ represents the lowest vertical wavenumber among the three waves at that particular stratification. Rotational effects are neglected ($f=0$).}
\label{fig:mismatchnvariation}
\end{figure}

To study the variation of detuning, four different values of $\omega_3/N_b$ are chosen where $\omega_3$ is the parent wave's angular frequency. $N_b$ is the background stratification where the vertical wavenumer triad condition is satisfied without any detuning ($\Delta m=0$). The background stratification where the vertical wavenumber condition is satisfied without any detuning is also referred as base stratification.
For each value of $\omega_3/N_b$, 
we consider four different combinations of daughter waves' angular frequencies. The daughter waves' respective angular frequencies are chosen by a parameter $\alpha$ such that $\omega_1 = (1-\alpha)\omega_3$  and $\omega_2 = \alpha\omega_3$. For each $\alpha$, there are four unique wavevectors for the daughter waves. The four unique triad combinations (for a particular $\alpha$ and $\omega_3/N_b$) can be characterized as:
\begin{enumerate}[label=(\alph*)\,\, ]
    \item {$(k_1/k_3,|{m_1/m_2}|)\in (1,\infty)\times (0,1)$}, 
    \item {$(k_1/k_3,|{m_1/m_2}|)\in (0,1)\times (1,\infty)$},
    \item {$(k_1/k_3,|{m_1/m_2}|)\in (0,1)\times (0,1)$},
    \item {$(k_1/k_3,|{m_1/m_2}|)\in (1,\infty)\times (1,\infty)$}.
\end{enumerate}
\noindent
Initially we study the effect of the variation of $\omega_3/N_b$ and $\alpha$ on detuning. To this end,  we focus on those triads whose daughter waves have wavenumbers satisfying  {$(k_1/k_3,|{m_1/m_2}|)\in (0,1)\times (1,\infty)$}. The results are given in figure \ref{fig:mismatchnvariation}. It can be observed that, for a given $\alpha$, detuning significantly increases as $\omega_3/N_b$ is increased for the same increase in the background stratification. Detuning asymptotes to a constant value as $N$ is increased, hence the difference in detuning caused by moderate and strong stratifications would be minimal. A given triad  satisfies resonant condition when $\Delta m=0$, which would occur only for a particular $N$; as the triads move to a different stratification (i.e. moving along a curve $\alpha=\textrm{constant}$), depending on $\omega_3/N_b$, the detuning effect could be small or large. We observe that detuning has a strong sensitivity to stratification for higher cases of $\omega_3/N_b$ values. For example, figure \ref{fig:mismatchnvariation}(a) shows that a  small variation in stratification causes  significant detuning for $\omega_3/N_b = 0.9$ near $N_b$. This effect purely arises from the dispersion relation of internal gravity waves.}

In figure \ref{fig:mismatchnvariation_differnt_triad}, we focus on the detuning for different wavevector (of daughter waves) combinations with $\omega_3/N_b$  fixed at $0.9$. We observe that out of all combinations, the triads satisfying {$(k_1/k_3,|{m_1/m_2}|)\in (1,\infty)\times (1,\infty)$} undergo the least amount of detuning with changes in the background stratification. Therefore, such triads may be the pathway through which the parent wave decomposes for high values of $\omega_3/N_b$. The triads shown in figure \ref{fig:mismatchnvariation_differnt_triad}(a) have values of non-dimensional detuning close to 1, which would mean that such triads are not possible in varying stratifications. Moreover, for a particular $\omega_3/N_b$ value, detuning can increase or decrease with an increase in $\alpha$ depending on the wavevector of the daughter waves. For example, figures \ref{fig:mismatchnvariation_differnt_triad}(a) and \ref{fig:mismatchnvariation_differnt_triad}(b) show that detuning increases with decrease in $\alpha$. However, for the triads in figures \ref{fig:mismatchnvariation_differnt_triad}(c) and \ref{fig:mismatchnvariation_differnt_triad}(d) detuning increases with an increase in $\alpha$.


\subsubsection{Nonlinear coupling coefficients} \label{section:5.1.3}
The nonlinear coupling coefficients ($\mathfrak{N}_j$) are functions of the vertical wavenumbers. Hence as the vertical wavenumber changes (when wave-packet moves to a different stratification), the nonlinear coupling coefficients will also change. The magnitude of nonlinear coupling coefficients (of all three waves) always increases (decreases) when the stratification increases (decreases) for waves which have angular frequency such that $\omega \ll N$ (which is shown in \eqref{eqn:NL_approx1}). 
This is consistent for any combination of subharmonic daughter waves. However, waves with $\omega \approx N$ do not have such monotonic increase (or decrease) for all possible subharmonic daughter waves. In such cases, whether the nonlinear coupling coefficients increase or decrease depend on the specific daughter wave combination. The nonlinear coupling coefficients are effectively proportional to the square root of the local stratification value for waves which have $\omega\ll N$, as shown below.

When $\omega\ll N$, the expression for the vertical wavenumber (given by \eqref{eqn:vertical_wavenumber}) can be approximated as:
\begin{equation}
    m_j = k_j \sqrt{\frac{N^2-\omega^2_j}{\omega^2_j-f^2}} \approx N \sqrt{\frac{k_j^2}{\omega^2_j-f^2}} = \zeta_jN,
 \label{eqn:vertical_approximation}
\end{equation}
where $\zeta_j \equiv \sqrt{{k_j^2}/{(\omega^2_j-f^2)}}$ is defined for convenience. It can be noticed that $\zeta_{j}$ does not change with stratification. In a similar way another approximation can be made:
\begin{equation}
    m_j^2 + k_j^2 = k_j^2 \left({\frac{N^2-f^2}{\omega^2_j-f^2}}\right) \approx k_j^2 {\frac{N^2}{\omega^2_j-f^2}}.
 \label{eqn:vertical_approximation_2}
\end{equation}
Now we use (\ref{eqn:vertical_approximation}) and (\ref{eqn:vertical_approximation_2}) in (\ref{eqn:coupling_coefficient_1}), resulting in (after simplification):
\begin{align}
\mathfrak{N}_{1}(\epsilon_{n} z) = &
\sqrt{N} \Tilde{\mathfrak{N}}_1,\label{eqn:NL_approx1}\\
\mathrm{where}\,\,\Tilde{\mathfrak{N}}_{1} &= \left[\frac{(k_{3}-k_{2})(\omega_1^2-f^2)}{k^{2}_{1}\omega_{1}}\left( \frac{k_{3}k_{2}\zeta_{2}}{\omega_{2}} + \frac{k_{3}k_{2}\zeta_{3}}{\omega_{3}}  - \frac{k_{3}^{2}\zeta_{2}}{\omega_{3}} - \frac{k_{2}^{2}\zeta_{3}}{\omega_{2}} \right)\left( \frac{\zeta_{1}}{{\zeta_{2}\zeta_{3}}} \right)^{1/2}\right]  \nonumber \\ 
        &-\left[\frac{(\omega_{1}^2-f^2)}{k^{2}_{1}}(k_{3}\zeta_{2}-k_{2}\zeta_{3})\left(  {\frac{k_3^2}{\omega^2_3-f^2}}-  {\frac{k_2^2}{\omega^2_2-f^2}}\right) \left( \frac{\zeta_{1}}{{\zeta_{2}\zeta_{3}}} \right)^{1/2}\right]  \nonumber \\
           &-\left[\frac{f^2(\zeta_3-\zeta_2)(\omega_1^2-f^2)}{k^{2}_{1}\omega_{1}} \left(\frac{\zeta_3 \zeta_2 k_3}{\omega_3} + \frac{\zeta_3 \zeta_2 k_2}{\omega_2} - \frac{\zeta_3^2k_2}{\omega_3} - \frac{\zeta_2^2k_3}{\omega_2} \right)\left( \frac{\zeta_1}{{\zeta_{2}\zeta_{3}}} \right)^{1/2} \right].
\label{eqn:NL_approx}
\end{align}
 Notice that $\tilde{\mathfrak{N}}_{j}$ does not change with stratification. A similar analysis can also be done for the other coupling coefficients which would yield  a similar result. 
 
 Here we summarize the key observations of \S
 \ref{section:5.1.1} to \S \ref{section:5.1.3}:
\begin{itemize}
    \item Wave with angular frequencies  $\omega \ll N$, the nonlinear coupling coefficients always decrease (increase) when the wave-packets move to a lower (higher) stratification.  For waves with  $\omega \approx N$, whether the nonlinear coupling coefficient increases or decreases depends on the daughter waves.
    \item The group speed of any wave-packet decreases (increases) as the packet moves to a region of higher (lower) stratification.
    \item The width (or length scale) of any wave-packet decreases (increases) as the packet moves to a region of higher (lower) stratification.
\end{itemize}

\subsubsection{Numerical experiments \label{section:5.1.4}}

\paragraph{a) Wave-packets satisfying $\omega \ll N$:}

\vspace{0.2cm}

Here we validate the theoretical layout given in \S \ref{section:5.1.1} to \S \ref{section:5.1.3} with numerical experiments. Initially we focus on waves with angular frequencies such that $\omega \ll N$. The governing equations (\ref{eqn:wave1})--(\ref{eqn:wave3}) are used in the inviscid limit with  $x$-independent amplitudes. The three evolution equations are solved using the same numerical procedure mentioned in \S \ref{Section:4}.


A triad having the following angular frequencies is chosen:
$\omega_{1} = 0.0375N_{b}$, $\omega_{2} = 0.0125N_{b}$, and $\omega_{3} = 0.05N_{b}$, where $N_b$ (chosen to be $10^{-3} \textnormal{s}^{-1}$) is the base stratification where the resonant triad condition $(\omega_1,k_1,m_1)+(\omega_2,k_2,m_2)=(\omega_3,k_3,m_3)$ is perfectly satisfied. The angular frequencies of the constituent waves are chosen such that  $\omega_j\ll N$. The horizontal wavenumbers are $k_{1}H = 1.12$, $k_{2}H = -0.12$, and $k_{3}H = 1$ (with $H = 1000\textnormal{m}$), and satisfy the resonant triad condition $k_1+k_2=k_3$. Rotational effects are ignored for simplicity ($f=0$). 

\begin{figure}
{\centering
\includegraphics[width=1.0\linewidth,keepaspectratio]{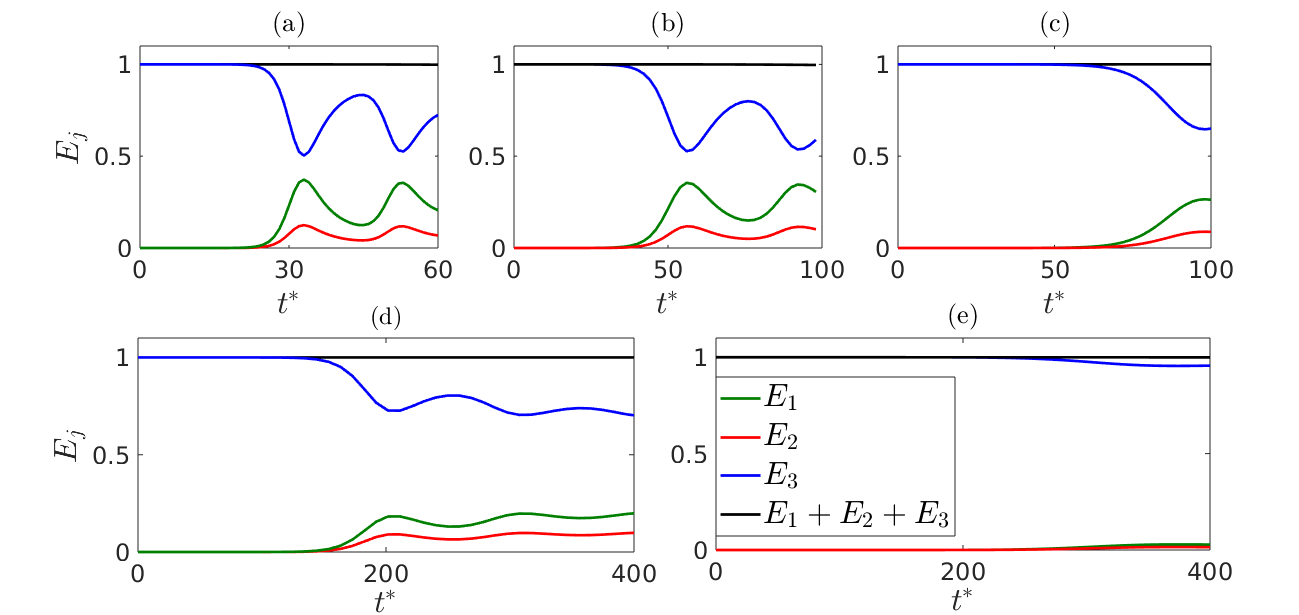}}
\caption[Energy evolution plot for triad case 4]{{ Comparison of time evolution of energy for wave-packet triads in uniform and weakly varying stratifications. In all cases, the waves form a resonant triad (triad conditions are perfectly met) at the base stratification $N_b$. Two different triads are considered, one for 
Cases 1--3 (sub-figures (a)--(c)), and another for Cases 4--5 (sub-figures (d)--(e)).
(a) Case 1:  Wave-packets moving to a region of higher stratification ($4N_b$) from the base stratification. 
(b) Case 2: Wave-packets interacting in a domain of constant stratification ($N_b$). The leads to slower energy transfer than Case 1. 
(c) Case 3: Wave-packets moving to a region of lower stratification ($0.4N_b$) from the base stratification. The energy transfer is slower than Case 2. (d) Case 4: Wave-packets interacting in a domain of constant stratification ($N_b$). Note this is a different triad than Case 2. (e) Case 5: Wave-packets interacting in a constant stratification ($1.04N_b$), which is only $4\%$ higher than the base stratification  $N_b$ for which the triad conditions are perfectly met. This small change causes very high reduction in the energy transfer.
}}
\label{fig:nonvsuniform60}
\end{figure}

Using the above-mentioned triad, three simulations are run in three different background stratifications. We consider the initial amplitude profile for all three wave-packets forming the triad to have a Gaussian distribution in $z$-direction. Therefore the amplitude definitions (\ref{eqn:amp_def_gauss_1}) are used, with  $A_1 = A_2 = 10^{-5}$ $\textnormal{m}^{5/2}\textnormal{s}^{-1}$, and $A_3 = 2 \times 10^{-2} $ $\textnormal{m}^{5/2}\textnormal{s}^{-1}$ for all simulations. 
The different stratifications used in the simulations are given below: 
\begin{enumerate}
\item Case 1 -- Wave-packets moved from $N_{b}$ (where resonant triad condition is perfectly met) to a new stratification region $4N_{b}$.
\item Case 2 -- Base stratification is held constant at $N_{b}$ throughout the domain.
\item Case 3 -- Wave-packets moved from $N_{b}$ (where resonant triad condition is perfectly met) to a new stratification region $0.4N_{b}$.
\end{enumerate}
The width of the wave-packets in all the three simulations are chosen according to the background stratification where the wave-packets have moved to. As mentioned in \S \ref{section:5.1.1}, wave-packets' width varies when they move to a region of a different stratification. 
Since the stratification remains constant at $N_{b}$ in Case 2, so remains the width of the wave-packets.
However, the issue of varying wave-packet width comes into play in Cases 1 and 3. In both cases, we assume that the wave-packets initially (i.e. when they are at $N=N_b$) have the same width as that in Case 2. The width at a later time when they move to a new stratification region ($4N_{b}$ in Case 1 and $0.4N_{b}$ in Case 3) can be found using (\ref{eqn:amp_fin}). Hence we finally obtain: 
\begin{enumerate}
\item Case 1 --  $W_{p(1)}=W_{p(2)}=W_{p(3)}=20\lambda_3$. 
\item Case 2 --  $W_{p(1)}=W_{p(2)}=W_{p(3)}=80\lambda_3$.
\item Case 3 --  $W_{p(1)}=W_{p(2)}=W_{p(3)}=200\lambda_3$,
\end{enumerate}
where $\lambda_{3}$ is the vertical wavelength of `wave-3' in stratification $N_b$.
{Interestingly, all the wave-packets' size  achieve a new constant value (different from that at $N=N_b$) for Cases 1 and 3 since the group-speed (the main determiner of wave-packet size) follows a simple inverse relationship with local stratification (as mentioned in \S \ref{section:5.1.1}).}
The simulation results are given in figure \ref{fig:nonvsuniform60}; surprisingly, energy transferred in the high buoyancy frequency region (Case 1, which does not satisfy the resonant triad condition and hence there is a detuning in the vertical wavenumber) is slightly higher in comparison to that in the uniformly stratified region (Case 2, where resonant condition is always met), compare figures \ref{fig:nonvsuniform60}(a) and \ref{fig:nonvsuniform60}(b). More importantly, we also observe that the energy is transferred more quickly from the parent wave in Case 1 than that in Case 2. Although detuning is present in Case 1, its effect is negligible since $\Pi_m \approx 480$. The rate of energy transfer is higher because the nonlinear coupling coefficients increase as $\sqrt{N}$; see \eqref{eqn:NL_approx1}. 
Increase in the growth rates in the higher stratification region (as observed in Case 1) was also reported in \cite{gayen},
where an internal wave beam propagates from a lower uniform stratification into a higher stratification region (pycnocline) and undergoes PSI inside the pycnocline (beam's frequency also remains constant in varying stratification, similar to our case).


In situations where wave-packets move to a region of lower stratification, in addition to wave detuning, this results in reduced nonlinear coupling coefficients, hence the growth rate of the daughter wave-packets will \emph{always} be lesser. This is what happens in Case 3, and is shown in figure \ref{fig:nonvsuniform60}(c). We note here that this particular case is only applicable for $\omega \ll N$. For waves with $\omega \approx N$, the nonlinear coefficients may increase or decrease with decrease in stratification.

An important point to note is that even though the energy transfer is increased in Case 1, it may not always be true for waves satisfying $\omega\ll N$. For example, wave-packets satisfying $\Pi_w \sim \mathcal{O}(1)$ and $\Pi_m \ll \mathcal{O}(1)$ (see \S \ref{Section:4}) may do the reverse - energy transfer may be lower than that for resonant triads (as is expected for non-resonant triads). This is due to the fact that, even though increase in stratification increases the nonlinear coupling coefficients, the presence of high detuning $\Pi_m \ll \mathcal{O}(1)$ (as a result of increasing the stratification) would render the wave-packets unable to exchange energy amongst themselves.  
Hence in summary, the growth rates of the daughter waves may increase or decrease when they move to a region of higher stratification (where the resonant condition is not satisfied) from a base stratification where the resonant condition is satisfied. The increase or decrease depends on the relative strengths of the group speed term and the nonlinear forcing term.

\vspace{0.2cm}

\paragraph{b) Wave-packets satisfying $\omega \approx  N$:}  

\vspace{0.2cm}

For wave-packet triads  satisfying $\omega\approx N$, even a small increase (or decrease) in stratification results in a high detuning of vertical wavenumbers for certain triads as shown in \S \ref{section:5.1.2}. However, a small change in stratification has nearly no effect on the group speeds and nonlinear coupling coefficients. Hence this detuning may reduce the energy transfer between the wave-packets.
To show this, we have performed two numerical simulations - one with a (constant) base stratification $N_b$, where the triad conditions are perfectly met throughout the domain (like Case 2 for $\omega\ll N$), and the other where the wave-packets move to a higher stratification region
(like Case 1 for $\omega\ll N$). However, the higher stratification region in this case is only slightly (4$\%$) higher than $N_b$ (the region where the triad condition is perfectly met). 


For these numerical experiments, we choose the frequencies of the constituent waves of the triad to be respectively  $\omega_{1} = 0.3N_b$, $\omega_{2} = 0.6N_b$, and $\omega_{3} = 0.9N_b$,  while the respective horizontal wavenumbers are $k_{1}H = -0.25$, $k_{2}H = -0.38$, and $k_{3}H = -0.63$. They satisfy the triad condition $(k_1,\omega_1) + (k_2,\omega_2) = (k_3,\omega_3)$. 
Furthermore, $N_b = 10^{-3} \textnormal{s}^{-1}$, $f=0$ and  $H = 100 \textnormal{m}$.
We consider the initial amplitude profile for all the three wave-packets forming the triad to be a Gaussian distribution in the $z$-direction. Therefore the amplitude definitions (\ref{eqn:amp_def_gauss_1}) are used, where $W_{p(1)} = W_{p(2)} = W_{p(3)} = 30\lambda_3$, $A_1 = A_2 = 10^{-5} $ $\textnormal{m}^{5/2}\textnormal{s}^{-1}$, and $A_3 = 6 \times 10^{-2} $ $\textnormal{m}^{5/2}\textnormal{s}^{-1}$.

The stratifications for the two simulations are given below: 
\begin{enumerate}
\item Case 4 -- Base stratification is held constant at $N_{b}$ throughout the domain.
\item Case 5 -- Wave-packets moved from $N_{b}$ (where resonant triad condition is perfectly met) to a new stratification region $1.04N_{b}$.
\end{enumerate}
There is no detuning in Case 4, but for Case 5, $\Delta m/m_{3} = 0.15$, which results in $\Pi_m = 0.51$.
Figure \ref{fig:nonvsuniform60}(d) shows the energy evolution in Case 4 and figure \ref{fig:nonvsuniform60}(e) shows the same for Case 5.
The rate of energy transfer for Case 4 is significantly more than that in Case 5, even though for the latter, the wave-packets interact in a region with 4$\%$ higher stratification. We re-emphasize that the energy transfer in this case is almost exclusively dictated by the detuning, nonlinear coupling coefficients and group-speeds are almost unaltered (for this particular example) by the slight change in stratification. The system here behaves like the case of $W_{p(1)} = 60\lambda_3$ in \S \ref{Section:4}.


It is important to note that, even though the energy transfer is significantly reduced for the slightly higher stratification when $\omega_j \approx N$, it may  not always be the case. In triad systems where the group speed term is much lesser than the nonlinear forcing term, even significant increase in detuning ($\Delta m$) may not reduce the energy transfer rates. That is, in systems where $\Pi_m \ll \mathcal{O}(1)$,  the effect of detuning is negligible. The growth rate in such systems can even increase if the nonlinear coupling coefficients increased (the increase would be very small in this case) as a result of a very small increase in the background stratification.

{\subsubsection{Effect of rotation}
Even though in all our analysis in \S \ref{section:5.1.4} the rotational effects were neglected, the results would be qualitatively similar if we included the rotational effects. For waves which have frequency such that $\omega \ll N$, we observe from expression (\ref{eqn:vertical_approximation}) that even with $f \neq 0$, the vertical wavenumber approximately becomes a linear function of stratification. This again leads to the group speeds of the waves being inversely proportional to the background stratification (shown in \eqref{eqn:GS_explain}) and the nonlinear coupling coefficients being proportional to the square-root of the stratification (shown in \eqref{eqn:NL_approx1}). Moreover, for such waves, this introduces little mismatch in the vertical wavenumber in comparison to waves which satisfies the condition $\omega \approx N$. Hence, we expect to find the same qualitative results as $f=0$ if the slow, $\mathcal{O}(1)$ variations in stratification are studied for a fixed $\omega$ and non-zero $f$. {Here we emphasize that the same results are also applicable for near-inertial parametric subharmonic instability where an internal gravity wave interacts with two daughter waves whose frequencies are almost the inertial frequency ($f$).}}

\subsection{Interacting wave-packets in weakly varying stratification: including viscous effects}
\label{section:5.2}
In this sub-section, we consider the effects of viscosity on the growth rates of the waves when they move to a region of different stratification. Here, the effect of viscosity is mainly 
considered for triads satisfying $\omega_3 \ll N$, 
where $\omega_3$ is the angular frequency of the parent wave. We consider the governing triad interaction equations \eqref{eqn:wave1}--\eqref{eqn:wave3} with an assumption of no $x-$dependence.

We analyze the growth rates of these waves when they travel to a constant stratification region $N$, which is different from the stratification $N_b$ (also a constant) where the waves were initially located and satisfied the resonant condition. In the stratification region $N$, the waves will have constant detuning in vertical wavenumber ($
\Delta m$).
Moreover, we normalize the  wave amplitudes as follows: $\hat{a}_j \equiv a_j/\sqrt{\abs{m_{j(b)}}}$, where $m_{j(b)}$ is the vertical wavenumber at $N_b$. Substituting this definition in \eqref{eqn:wave1}--\eqref{eqn:wave3} yields
 \begin{subequations}
\begin{align}
\frac{\partial \hat{a}_{1}}{\partial t} + c_{z,1}^{(g)}\frac{\partial \hat{a}_{1}}{\partial z}   + \mathcal{V}_1\hat{a}_1 &= \frac{1}{2} \hat{ \mathfrak{N}}_{1}{\hat{a}}_{3}\bar{\hat{a}}_{2} \ee^{\ii \Delta mz},\label{eqn:vis_reduced1_hat}\\
\frac{\partial \hat{a}_{2}}{\partial t} + c_{z,2}^{(g)}\frac{\partial \hat{a}_{2}}{\partial z}  + \mathcal{V}_2\hat{a}_2  &= \frac{1}{2} \hat{ \mathfrak{N}}_{2}{\hat{a}}_{3}\bar{\hat{a}}_{1} \ee^{\ii \Delta mz}, \label{eqn:vis_reduced2_hat} \\
\frac{\partial \hat{a}_{3}}{\partial t} + c_{z,3}^{(g)}\frac{\partial \hat{a}_{3}}{\partial z}  + \mathcal{V}_3\hat{a}_3  &= \frac{1}{2} \hat{ \mathfrak{N}}_{3}{\hat{a}}_{1}{\hat{a}}_{2} \ee^{-\ii \Delta mz}.
\label{eqn:vis_reduced3_hat} 
\end{align}
\end{subequations}
Here $\hat{\mathfrak{N}}_j \equiv \mathfrak{N}_{j}\beta_j$, and $\beta_j$ is defined as:
\begin{equation}
 \beta_1 \equiv  \sqrt{\abs{\frac{m_{2(b)}m_{3(b)}}{m_{1(b)}}}}, \hspace{0.5cm}  \beta_2 \equiv  \sqrt{\abs{\frac{m_{3(b)}m_{1(b)}}{m_{2(b)}}}}, \hspace{0.5cm}
 \beta_3 \equiv  \sqrt{\abs{\frac{m_{1(b)}m_{2(b)}}{m_{3(b)}}}}.
\end{equation}
To estimate the growth rate of the daughter waves, we first assume  the parent wave has several orders of magnitude higher energy than the daughter waves.
Next we assume the parent wave's amplitude, $\hat{a}_{3}$, to remain constant in time and space so as to estimate the growth rates of the daughter waves (this behavior is expected in the early stages of growth of the daughter waves).
The assumption that $\hat{a}_{3}$ is constant in space is legitimate when the parent wave-packet width is considered to be large, and under these assumptions,  (\ref{eqn:vis_reduced3_hat}) becomes trivial and can therefore be ignored. We note in passing that the parent wave had arbitrary length scale in \S \ref{Section:3}, hence $\hat{a}_{3}$ was assumed to be of normal mode type (and not constant).

We assume normal mode form for amplitudes of the daughter waves: $\hat{a}_1 = \Tilde{a}_{1}(\epsilon_{t} t)\ee^{\ii (M_{1}z)}$ and $\hat{a}_1 = \Tilde{a}_{2}(\epsilon_{t} t)\ee^{\ii (M_{2}z)}$, where $M_1$ and $M_2$ are respectively the \emph{vertical wavenumbers of the amplitudes profiles} $a_1$ and $a_2$  (not to be confused with vertical wave numbers, $m_j$) in the stratification region $N$. Moreover, similar to \S \ref{Section:3}, we consider normal modes such that $M_{1}+M_{2} = \Delta m$, hence the governing equations can be reduced to a purely temporal form. Using all the aforementioned assumptions, the resulting evolution equations for the daughter waves are:
\begin{subequations}
\begin{align}
\frac{\partial \Tilde{a}_{1}}{\partial t} + \ii c_{z,1}^{(g)}M_1 \Tilde{a}_{1} + \mathcal{V}_1\Tilde{a}_1 &= \frac{1}{2} \hat{ \mathfrak{N}}_{1}A_3\bar{\Tilde{a}}_{2}, \label{eqn:vis_reduced1_time}\\
\frac{\partial \Tilde{a}_{2}}{\partial t} + \ii c_{z,2}^{(g)} M_2\Tilde{a}_{2} + \mathcal{V}_2\Tilde{a}_2  &= \frac{1}{2} \hat{ \mathfrak{N}}_{2}A_3\bar{\Tilde{a}}_{1}, \label{eqn:vis_reduced2_time} 
\end{align}
\end{subequations}
where $A_3$ is the parent wave amplitude. 
The solution for $\Tilde{a}_{1}$ in \eqref{eqn:vis_reduced1_time} and \eqref{eqn:vis_reduced2_time} can be found by assuming solutions of the form: $\Tilde{a}_{1} = A_{+}\exp{({\sigma}_{+}t)}+A_{-}\exp{({\sigma}_{-}t)}$, where $(A_{+}$ and $A_{-})$ are constants. The growth rates $\sigma_{\pm}$ obtained are as follows:


\begin{align}
   {\sigma}_{\pm} = &- \frac{1}{2}\left[\ii (\widehat{M}_1 - \widehat{M}_2) +\mathcal{V}_1 + \mathcal{V}_2\right] \pm \frac{1}{2} \sqrt{\left[\ii (\widehat{M}_1 + \widehat{M}_2) + \mathcal{V}_1 - \mathcal{V}_2\right]^2 + \hat{\mathfrak{N}}_{2} \hat{\mathfrak{N}}_{1}A_3^2} ,
\label{eqn:vis_CG_gr}
\end{align} 
where $\widehat{M}_j \equiv M_j c_{z,j}^{(g)}$ is defined for convenience. {An expression bearing resemblance with \eqref{eqn:vis_CG_gr}  was obtained in \cite{bourget_width_2014} and \cite{maurer_2016}. However their expression, derived using control volume analysis, is  limited to constant stratification.}
The growth rate of a particular normal mode is given by $\textnormal{Re}(\sigma_{\pm})$, where $\textnormal{Re}(\,)$ denotes the real part. The expression for $\textnormal{Re}(\sigma_{\pm})$ is cumbersome and thus avoided for brevity, however it is straight-forward to observe  that $\textnormal{Re}(\sigma_{\pm})$ does not contain $\widehat{M}_1 - \widehat{M}_2$. However, terms containing $\widehat{M}_1 + \widehat{M}_2$ do appear, and it is important to understand the significance of this term. Note that $\widehat{M}_1 + \widehat{M}_2= M_1 c_{z,1}^{(g)}+M_2 c_{z,2}^{(g)}$, i.e., it is a weighted  (by vertical group speed) sum of $M_j$s. Since the detuning  $\Delta m=M_{1}+M_{2}$ by assumption, this implies that the effect of detuning between the normal modes is captured only through this term. Monotonically increasing $\widehat{M}_1 + \widehat{M}_2$  decreases the growth rate regardless of the viscosity. In order to single out the effect of viscosity on the growth rates, the parameters are chosen such that the effects of detuning can be neglected (see the scaling analysis \eqref{eqn:scaling_analysis_1}), i.e. we consider $\mathcal{O}(\widehat{M}_1 + \widehat{M}_2) \ll \mathcal{O}(\mathcal{V}_1,\mathcal{V}_2)$.
 Hence the growth rate expression \eqref{eqn:vis_CG_gr} can be simplified to:
\begin{equation}
{\sigma}_{\pm} = - \frac{1}{2} \left(\mathcal{V}_1 + \mathcal{V}_2\right)\pm \frac{1}{2}\sqrt{\left(\mathcal{V}_1 - \mathcal{V}_2\right)^2 + \hat{\mathfrak{N}}_{2} \hat{\mathfrak{N}}_{1}A_3^2}. 
\label{eqn:vis_gr_simple}
\end{equation}
Before parametrically exploring (\ref{eqn:vis_gr_simple}), the assumptions (\ref{eqn:vertical_approximation}), (\ref{eqn:vertical_approximation_2}) and (\ref{eqn:NL_approx1}) are used to simplify the growth rate expression (\ref{eqn:vis_gr_simple}), which is as follows:
\begin{equation}
    {\sigma}_{+} = -N^2\frac{\nu}{4}\left[\widehat{\zeta}_1^2+\widehat{\zeta}_2^2\right] + \sqrt{N^4\left[\frac{\nu^2}{16}\left(\widehat{\zeta}_1^2-\widehat{\zeta}_2^2\right)^2\right] + N\left[\Tilde{\mathfrak{N}}_{1}\Tilde{\mathfrak{N}}_{2}\beta_1\beta_2A_{3}^2\right]}.
\label{eqn:vis_gr_simplified}
\end{equation}
where $\widehat{\zeta}_j^2 = \zeta_j^2\left(1+{f^2}/{\omega_j^2}\right)$ is defined for convenience. For simplicity, from now on we drop the `+' sign in $\sigma_{+}$. We observe that the viscous terms are proportional to $N^{2}$ where as the nonlinear coupling term is proportional to $\sqrt{N}$. Hence, an increase in stratification may reduce or increase the growth rate of the daughter waves depending on the strength of the viscous term and the nonlinear coupling term.

Now we parametrically explore equation (\ref{eqn:vis_gr_simple}). For the analysis, the angular frequency of the parent wave is fixed at $\omega_3/N_b = 0.1$. The parent wave's wavenumbers are chosen such that they satisfy the dispersion relation dictated by the chosen $\omega_{3}/N_{b}$ value. The actual magnitude of the wavenumbers do not  qualitatively change the growth rates of the daughter waves  but only quantitatively for a given $\omega_3/N_b$  (provided the group speed term is not of the same order of magnitude of the viscous term or the nonlinear forcing term).
To classify the daughter waves, the parameter $\alpha$ (defined in \S \ref{section:5.1.2}) is used. Moreover, we also use the same classification used in \S \ref{section:5.1.2} for the different wave vectors possible for the same $\alpha$. The parameter $\alpha$ is varied, and for all the resulting triads, the change in growth rate with the change in stratification is studied by using the expression (\ref{eqn:vis_gr_simple}). A point to notice is that,  similar to \S \ref{section:5.1}, the stratification is varied without varying the angular frequencies of the waves. 

\begin{figure}
\centering
\includegraphics[width=1.0\linewidth]{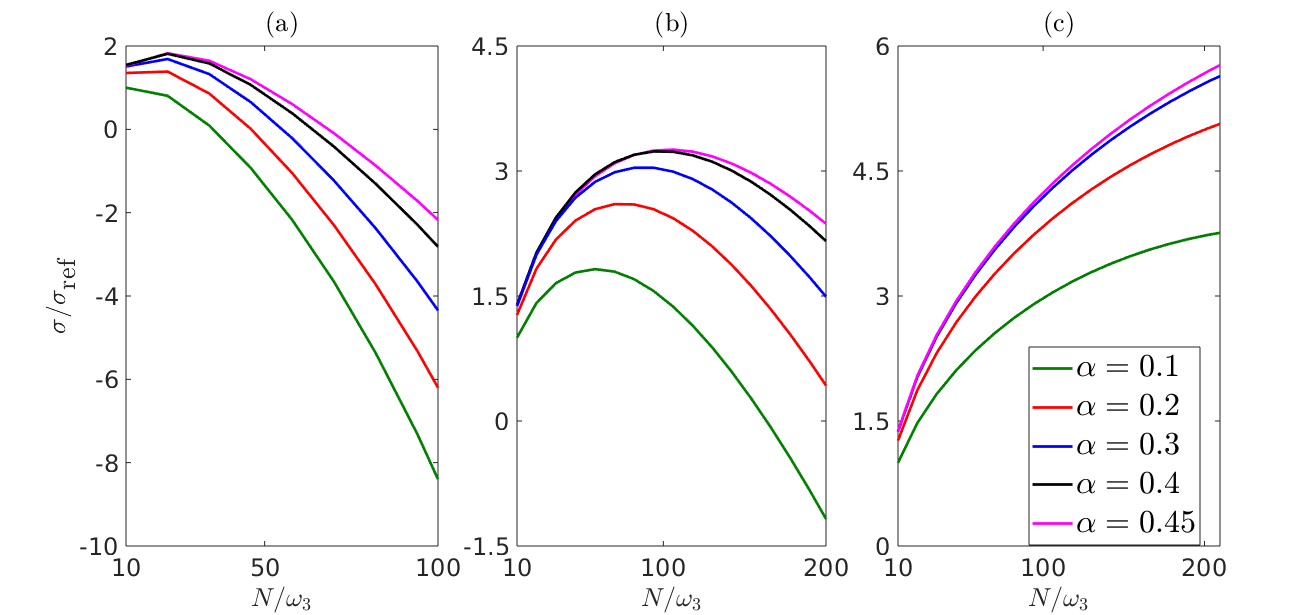}
\caption{ 
Growth rates versus stratification.  The  daughter wave combinations are chosen such that the horizontal wavenumbers satisfy {$(k_1/k_3,|{m_1/m_2}|)\in (0,1)\times (0,1)$}.  The ratio of the parent wave amplitude and kinematic viscosity  ($A_{3}/\nu$) are: (a) $A_3/\nu = 10^2$,  (b) $A_3/\nu = 10^3$, and  (c) $A_3/\nu = 10^4$.
}
\label{fig:k1lessk3}
 \end{figure}

\begin{figure}
\centering
\includegraphics[width=1.0\linewidth]{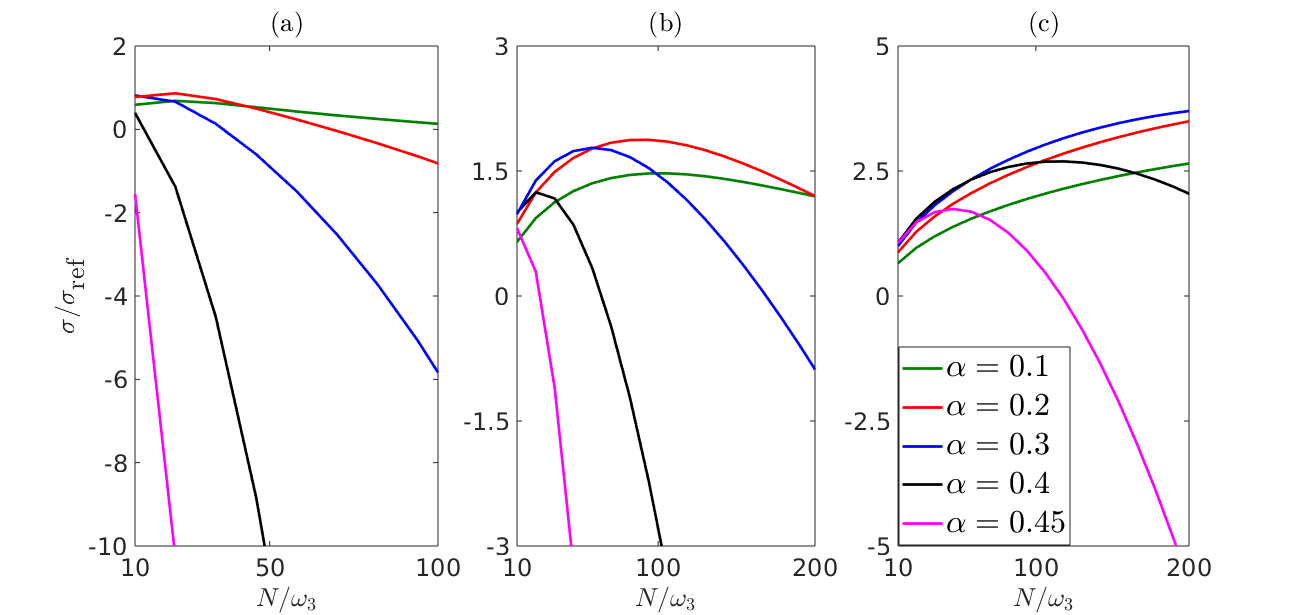}
\caption{Growth rates versus stratification.  The  daughter wave combinations are chosen such that the horizontal and vertical wavenumbers respectively satisfy {$(k_1/k_3,|{m_1/m_2}|)\in (1,\infty)\times (1,\infty)$}.  The ratio of the parent wave amplitude and kinematic viscosity  ($A_{3}/\nu$) are: (a) $A_3/\nu = 10^2$,  (b) $A_3/\nu = 10^3$, and  (c) $A_3/\nu = 10^4$. }
\label{fig:k1morek3}
 \end{figure}

 The ratio of kinematic viscosity and amplitude of the parent wave (note that this ratio is a non-dimensional quantity according to the definition of $\hat{a}_j$ given in this sub-section) are chosen to be: $A_{3}/\nu = 10^2, 10^3$ and $10^4$. 
  The variation of the growth rates for the various triads in the presence of viscosity are shown in figures \ref{fig:k1lessk3} and \ref{fig:k1morek3}.
The growth rates of all the triads are non-dimensionalized with a particular reference value of growth rate ($\sigma_{\textnormal{ref}}$), which occurs for the triad characterised by: $\alpha = 0.1$ for the triad {$(k_1/k_3,|{m_1/m_2}|)\in (0,1)\times (0,1)$} and $A_{3}/\nu = 10^2$ at base stratification $N_b$.  

We observe from figure \ref{fig:k1lessk3} that at lower values of $A_3/\nu$ (figure \ref{fig:k1lessk3}(a)), increase in the stratification can reduce the growth rates of the daughter waves significantly. This is true for all $\alpha$ values. This is opposite to what was observed in the inviscid case, where the growth rate increased with an increase in stratification in the parameter regime $\omega_3 \ll N$, when detuning had negligible effect. For the case of $A_3/\nu = 10^{3}$,  as viscosity is increased the increase in stratification initially increases the growth rate and then again starts to decrease as the stratification is further increased; see figure \ref{fig:k1lessk3}(b). However for $A_3/\nu = 10^{4}$, figure \ref{fig:k1lessk3}(c) reveals that an increase in stratification simply increases the growth rate because the viscous term is too weak in comparison to the nonlinear resonant term which forces the daughter waves. Hence, as dictated by expression $(\ref{eqn:NL_approx1})$, the growth rate increases as a function of $\sqrt{N}$. The group of triads characterised by {$(k_1/k_3,|{m_1/m_2}|)\in (0,1)\times (1,\infty)$} is not given here since the behaviour is found to be qualitatively similar to {$(k_1/k_3,|{m_1/m_2}|)\in (0,1)\times (0,1)$} for all values of $A_{3}/\nu$.

The effect of increased stratification in growth rates of the daughter waves which have wavenumbers such that {$(k_1/k_3,|{m_1/m_2}|)\in (1,\infty)\times (1,\infty)$} is shown in figure \ref{fig:k1morek3}(a)-(c). We observe that the growth rates are more rapidly reduced with an increase in stratification in comparison to figure \ref{fig:k1lessk3} for all $\alpha$ values. For higher values of $\alpha$ (such as $\alpha = 0.4$ and $0.45$), the growth rate is reduced even for $A_3/\nu = 10^{4}$ for the given range of $N$; see figure \ref{fig:k1morek3}(c). This is because in this group of triads, the daughter waves have higher wavenumbers (in comparison to group of triads which are characterised by {$(k_1/k_3,|{m_1/m_2}|)\in (0,1)\times (0,1)$} and {$(k_1/k_3,|{m_1/m_2}|)\in (0,1)\times (1,\infty)$}) for the same parent wave which results in viscous terms being significantly large.
The other combination of triad such that {$(k_1/k_3,|{m_1/m_2}|)\in (1,\infty)\times (0,1)$} is not shown here because the growth rates of the triads at the base stratification $(N_{b})$ itself were much smaller in comparison to other two combinations of triads shown in figures \ref{fig:k1lessk3} and \ref{fig:k1morek3}. Moreover, increasing stratification caused a rapid decrease in the growth rates for all values of $A_3/\nu$. This is because these group of triads have much lesser values of $\mathfrak{N}_j$ in comparison to the group triads shown in figures \ref{fig:k1lessk3} and \ref{fig:k1morek3}.
 
The above analysis was done without considering rotational effects. However, even adding rotational effects qualitatively produced the same results for the same values of $A_3/\nu$. Furthermore, varying $\omega_3/N_b$ from $0.1$ to $0.05$ showed a qualitatively similar picture. However if $\omega_3 \approx N_b$,  change in the background stratification may cause significant detuning, leading to a violation of the triad condition (`near resonance' would not be applicable anymore), and is therefore not particularly suitable for this analysis.



\subsection{Determining the optimal base stratification
}
\label{sec:optimal_base}
Up to this point we have fixed the background stratification where the resonant condition is perfectly met (specified as the base stratification $N_b$). Here we intend to determine the \emph{optimal} base stratification that causes maximum energy transfer among the wave-packets {in a medium of varying} background stratification region. 

For this analysis, we assume a medium where the background stratification varies from $N_{\textnormal{L}}$ to $N_{\textnormal{H}}$, and the ratio  $N_{\textnormal{H}}/N_{\textnormal{L}}$ is varied. We assume $N_{\textnormal{H}}>N_{\textnormal{L}}$, and fix the parent wave's frequency: $\omega_3/N_{\textnormal{L}} = 0.1$ (satisfying the regime  $\omega_3 \ll N$)\footnote{We also studied other values of $\omega_3/N_{\textnormal{L}}$ respecting $\omega_3 \ll N$ and found results quite similar  to that reported in this subsection.}. The analysis has been performed under inviscid conditions.
First we study the case  $N_{\textnormal{H}}/N_{\textnormal{L}} = 10$.

For simplicity, we also assume that the parent wave-packet's amplitude ($a_3$) to be invariant in space and time; therefore $a_3$ remains constant even when the parent wave-packet propagates through varying stratification. Let us now consider a situation where the parent wave-packet is a part of two separate triads. Hence there are two separate daughter wave-packet duos forming a resonant triad with the given parent wave-packet. The first daughter wave-packet duo satisfies the resonant condition ($\Delta m=0$) with the parent wave at $N=N_{\textnormal{L}}$ (this triad is referred to as `triad $N_{\textnormal{L}}$'), while the second duo satisfied at $N=N_{\textnormal{H}}$ (referred to as `triad $N_{\textnormal{H}}$'). We characterize the daughter wave-packet duos by $\alpha = 0.25$,
where $\alpha$ has been  defined in \S \ref{section:5.1.2}. This means that the angular frequency $\omega_1$ of triad $N_{\textnormal{L}}$ is the same as that in triad $N_{\textnormal{H}}$, and the same condition holds for $\omega_2$.
The horizontal wavenumbers of the daughter waves in `triad $N_{\textnormal{H}}$' are $k_1/k_3 = 0.9375$ and $k_2/k_3 = 0.0625$, and the same for `triad $N_{\textnormal{L}}$' are $k_1/k_3 = 0.9372$ and $k_2/k_3 = 0.0628$. 

\begin{figure}
{\centering
\includegraphics[width=1.0\linewidth,keepaspectratio]{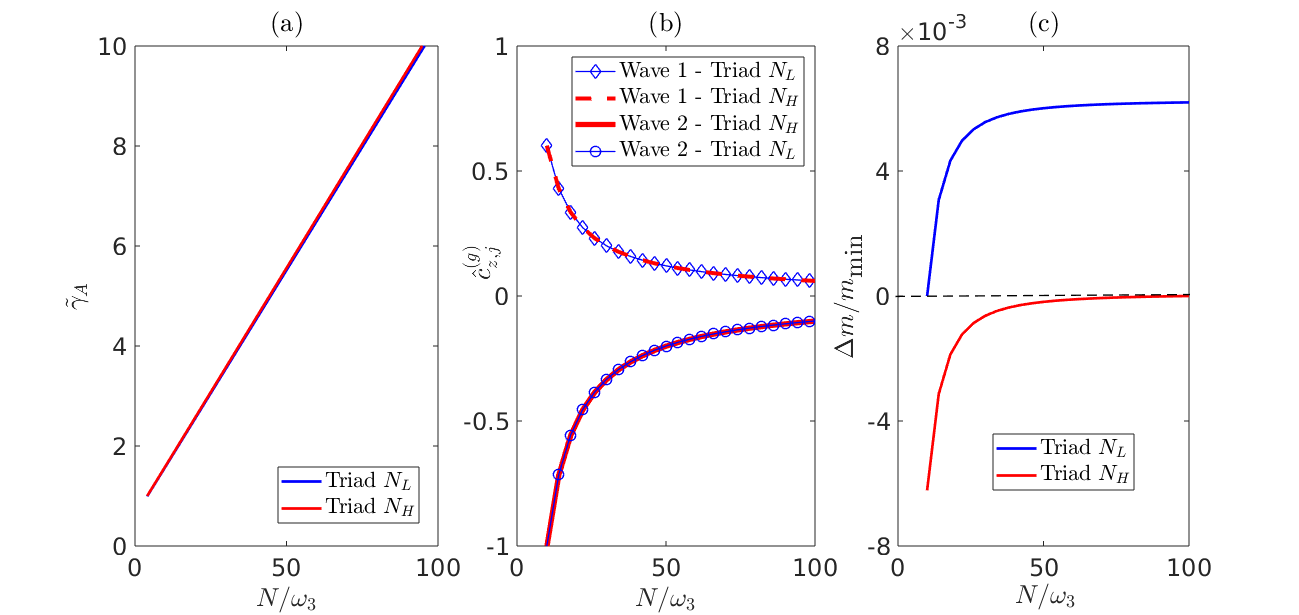}}
\caption[Energy evolution plot for triad case 4]{Comparisons between the `triad $N_{\textnormal{L}}$' and triad `$N_{\textnormal{H}}$' with varying stratification. (a) non-dimensional growth rate $\Tilde{\gamma}_A$, (b) non-dimensional group speeds $\hat{c}_{z,j}^{(g)}$ of the daughter waves, and (c) non-dimensional detuning $\Delta m/m_{\textnormal{min}}$.} \label{fig:selection}
\end{figure}

For both triads, the nonlinear resonant forcing term ${\gamma}_A$ (given in expression \eqref{eqn:gammaa_and_gammam}), the group speeds of the daughter waves, and the non-dimensional detuning $\Delta m/m_{\textnormal{min}}$ are plotted in figure \ref{fig:selection} as the background stratification is varied from $N_{\textnormal{L}}$ to $N_{\textnormal{H}}$. The quantity ${\gamma}_A$ is non-dimensionalized with $\gamma_A$ of the `triad $N_{\textnormal{L}}$' at the stratification $N_{{\textnormal{L}}}$. This non-dimensionlized $\gamma_A$ (given by $\Tilde{\gamma}_A$) serves as a measure of  the nonlinear forcing by the parent wave. Figure \ref{fig:selection}(a) shows that $\Tilde{\gamma}_A$ increases linearly for both triads, and are almost indistinguishable. This linear variation in $\Tilde{\gamma}_A$ with background stratification comes as follows. Combining \eqref{eqn:gammaa_and_gammam} and \eqref{eqn:NL_approx1} we obtain
\begin{equation*}
\gamma_{A}= \mathfrak{N}_{1}\mathfrak{N}_{2}A_{3}^{2} \propto N.
\end{equation*}

We also observe in figure \ref{fig:selection}(b) that the group speeds of the daughter wave-packets, $\hat{c}_{z,j}^{(g)}$, (group speed is non-dimensionalized by the parent wave's group speed at the stratification $N_{\textnormal{L}}$) for both the triads are nearly the same and almost indistinguishable. This near equality of group speeds, just like that observed for the growth rates, arise from nearly the same $k$ and $\omega$ values of the daughter wave-packet duos.
Furthermore, group speeds are found to follow an inverse law, which straight-forwardly comes from \eqref{eqn:GS_explain}:
\begin{equation*}
  c_{z,j}^{(g)} \propto \frac{1}{N}. 
\end{equation*}

While the nonlinear resonant forcing terms and group speeds of the two triads are nearly identical, the behavior of the vertical wavenumber detuning is non-trivial; see figure \ref{fig:selection}(c).

The detuning profile for the `triad $N_{\textnormal{L}}$' jumps from $0$ (no detuning) to its (near) maximum value in a short interval, and then asymptotes to the maximum value.
However  the reverse happens for the `triad $N_{\textnormal{H}}$' - detuning drops from its maximum magnitude to $0$ (no detuning) in a short interval. Hence `triad $N_{\textnormal{H}}$' \emph{stays} as a resonant triad for nearly the entire parameter space from $N_{\textnormal{L}}$ to $N_{\textnormal{H}}$. Therefore, `triad $N_{\textnormal{H}}$' is more conducive in transferring energy from the primary to the daughter waves in comparison to `triad $N_{\textnormal{L}}$'. We note here that although 
this entire study was for $\alpha = 0.25$,  similar qualitative behaviour was also observed for $\alpha = 0.05, 0.15, 0.35$, and $0.45$.


\subsection{Numerical validation of multiple scale analysis results} \label{section:5.3}
 
In this subsection we validate some results obtained from the reduced equations obtained via multiple scale analysis i.e. \eqref{eqn:wave1}--\eqref{eqn:wave3} with numerical simulations that solve the 2D Boussinesq Navier-Stokes equations. The equations \eqref{eqn:wave1}--\eqref{eqn:wave3} are numerically solved following the same procedure outlined in \S \ref{Section:4}.   
Similar to \S \ref{Section:4} and \S \ref{section:5.1}, we consider the initial amplitude profiles of all the three wave-packets forming the triad to be a Gaussian distribution in the $z$-direction. Therefore the amplitude definitions (\ref{eqn:amp_def_gauss_1}) are  used. We consider three validation cases:

\begin{itemize}
    \item \,\, Case 1: Here the amplitudes of the triad are given by $A_{1} = A_{2} = 0.5 \times 10^{-4} $, $A_{3} = 1 \times 10^{-2}$, $W_{p(1)} = W_{p(2)} = W_{p(3)} = 14\lambda_{3}$, where $\lambda_{3}$ is the vertical wavelength of wave-3. A uniform stratification $N_b= 10^{-3} \textnormal{s}^{-1}$ is considered; furthermore, $\omega_{1} = 0.124N_{b}$, $\omega_{2} = 0.0925N_{b}$,  $\omega_{3} = 0.216N_{b}$,  and $k_{1}H = -0.25$, $k_{2}H = -0.5$, $k_{3}H = -0.75$, where $H = 100 \textnormal{m}$. {Moreover, the vertical wavenumbers for this particular $N_b$ are: $m_1H=-2$, $m_2H=5.38$ and $m_3H=3.38$.}
We readily observe that the resonant condition $(k_1,m_1,\omega_1)+(k_2,m_2,\omega_2)=(k_3,m_3,\omega_3)$ is met. {The waves' amplitudes have been chosen such that they mimic a PSI-like situation.}
    \item \,\, Case 2: Here we consider the same triad as in Case 1, but in the background stratification $3N_b$. As a consequence of different background stratification, the vertical wavenumbers of the waves are different from Case 1.  
\item \,\, Case 3: {We also validate the reduced equations in the presence of a weakly varying stratification
\begin{equation}
 N = N_b + N_{\textnormal{max}}\exp[-(z/W_{n})^2],
 \label{eqn:varyingn}
 \end{equation}
where $N_b= 10^{-3} \textnormal{s}^{-1}$, $N_{\textnormal{max}}= 2N_b$ and $W_{n} = 8.5\lambda_{3}$. We use the same triad as Case 1 (i.e. same set of wavenumbers and frequencies), except the amplitudes and wave-packet widths are different; 
$A_{1} = A_{2} = -1\ii  \times 10^{-4} $, $A_{3} = - 0.5\ii \times 10^{-2}$, $W_{p(1)} = W_{p(2)} = W_{p(3)} = W_n$. }{We observe that the length scale of variation in $N$ and that of $a_j$ are the same in this case.}
\end{itemize}

The angular frequencies in  validation Cases 1--3 are chosen such that they satisfy $\omega\ll N_b$ for all three waves. The choice of the wavenumbers of the triads under consideration is primarily because the horizontal wavenumbers of wave-2 and wave-3 are integer multiples of wave-1, which would allow the periodic condition in $x$-direction to be enforced for a single wavelength of wave-1 (smallest wavenumber in the $x$-direction), thereby yielding a less expensive computation.
  \begin{figure}
{\centering
\includegraphics[width=1.0\linewidth,keepaspectratio]{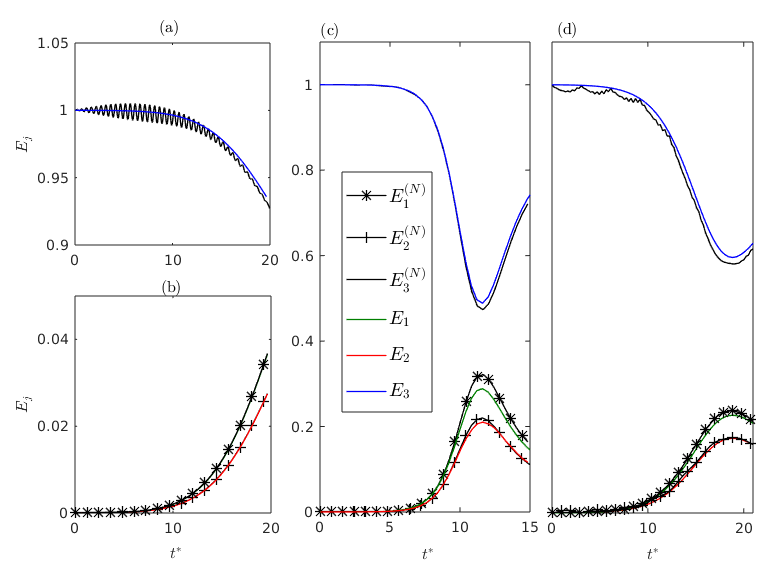}}
\caption[Energy evolution plot for triad case 4]{Comparison of numerically obtained energy transfer (denoted by superscript `$(N)$') with that obtained from multiple scale analysis. (a) Decay of the parent wave-packet in Case 1, and  (b) growth of the daughter wave-packets in Case 1. (c)  Growth of the daughter wave-packets and the decay of the parent wave-packet for a background stratification in Case 2. {(d) Energy evolution of the parent and daughter waves in Case 3.}
The non-dimensional time, $t^{*} = t\omega_3/2\pi$.}
\label{fig:validation}
\end{figure} 

The numerical validations are performed using an open source pseudo-spectral code \emph{Dedalus} \citep{dedalus} {-- the governing equations (\ref{eqn:NS_stream}) and (\ref{eqn:material_cons}) are solved with vanishing viscosity and $f=0$.}
{The problem is initialized with equivalent amplitude functions corresponding to the functions used in the multiple scale analysis. The equivalent amplitude functions in Dedalus are such that the initial velocity field of the waves in Dedalus and multiple scale formulation are the same.}
Both for Cases 1 and 2, we respectively consider $60$ and $4000$ Fourier modes  in $x$ and $z$ directions. 
{Moreover, we respectively consider $80$ and $1200$ Fourier modes  in $x$ and $z$ directions for Case 3.} 
Time marching is performed using semi-implicit backward differencing scheme, furthermore for time-stepping, 1500 steps per time-period of the parent wave is chosen for both Cases 1 and 2. {For Case 3, 1500 steps per time-period of the wave-1 is used.}
{ We have compared the potential energy of each wave obtained from multiple scale analysis with that obtained from Dedalus simulations. Since the total energy is equipartitioned between potential and kinetic (in the absence of rotation), the total energy is simply twice of the potential energy. 
}

\begin{figure}
 {\centering
 \includegraphics[width=1.0\linewidth,keepaspectratio]{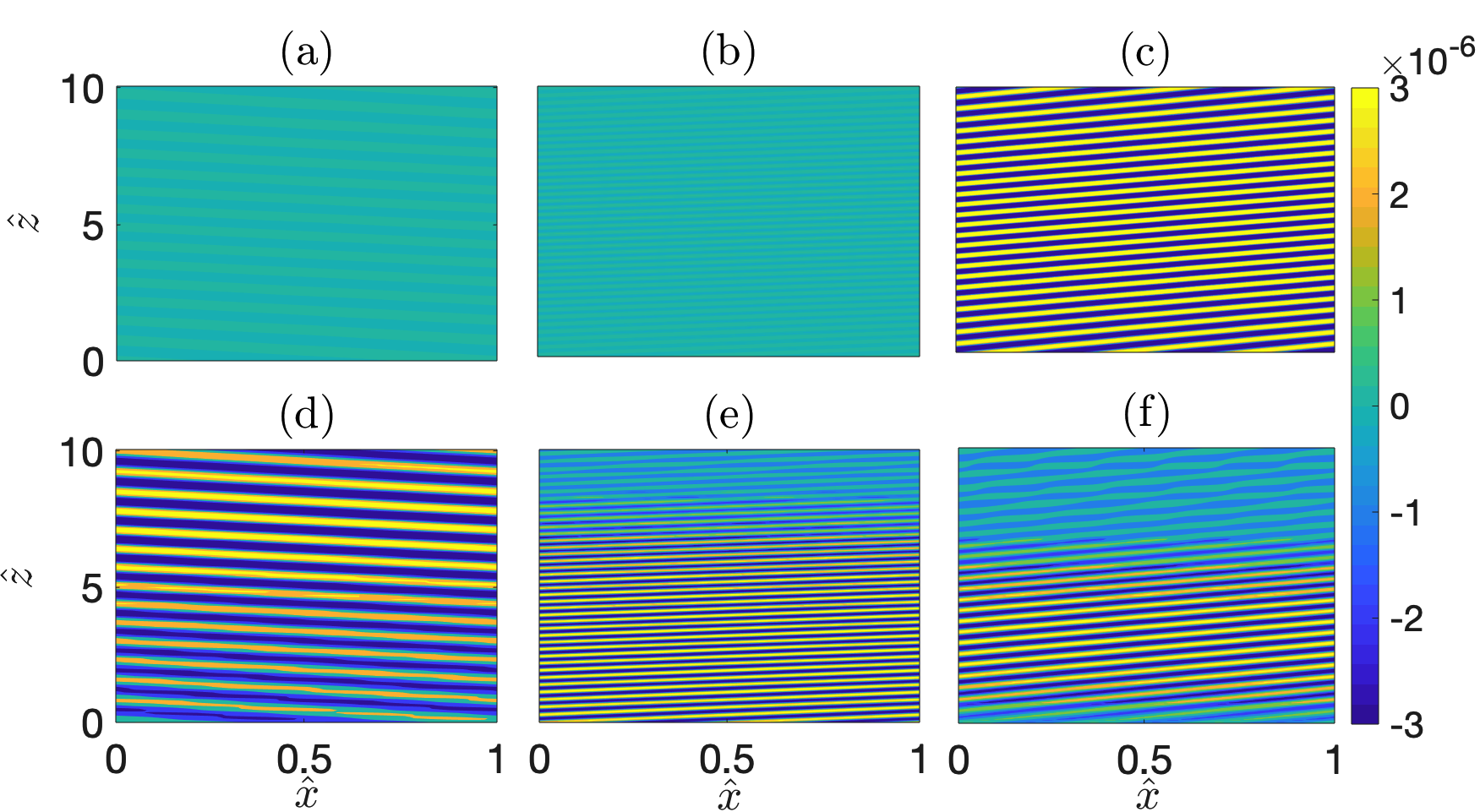}}
 \caption[Energy evolution plot for $\mathcal{O}(1)$ detuned systems.]{{ Contour plots of wave-filtered buoyancy perturbation field $(b)$. The plots are from Dedalus simulation of \S \ref{section:5.3}'s Case 3. (a), (b) and (c) respectively show buoyancy perturbation of wave-1, wave-2 and wave-3 at $t^{*} = 0$, while (d), (e) and (f) respectively show buoyancy perturbation of wave-1, wave-2 and wave-3 at $t^{*} = 21$ . Here  $\hat{x} = k_1x/2\pi$ and $\hat{z} = m_1z/2\pi$.}}
 \label{fig:contour}
 \end{figure}

Figures \ref{fig:validation}(a) and \ref{fig:validation}(b) respectively show the decay of the parent wave-packet and the growth of the daughter wave-packets  for Case 1. In figure \ref{fig:validation}(c), the growth of the daughter wave-packets and the decay of the parent wave-packet are shown when the stratification is  increased to $3N_b$, keeping the horizontal wavenumbers, angular frequencies and wave-packet sizes unchanged. However the vertical wavenumbers are dependent on the background stratification, and in this case we have: $m_1H= -6.04 $, $m_2H= 16.2$ and $m_3H= 10.38$. This implies that the resonant condition is not met - the wave triads are weakly detuned.
However, on increasing the stratification from $N_{b}$ to 3$N_{b}$,  we observe increased growth rates of the daughter wave-packets in a shorter time.
This is due to the fact that higher stratification increases the nonlinear coupling coefficients and reduces the group speed, which is in accordance with the findings in \S \ref{section:5.1}. 
{In figure \ref{fig:validation}(d), the energy evolution of the waves are shown for the case of varying background stratification (Case 3).  We observe that the numerical results match reasonably well with that of multiple scale analysis for all the three cases. In addition, a contour plot of the wave-filtered buoyancy perturbation corresponding to Case 3 is given in figure \ref{fig:contour}. }

{ \subsection{ Effect of variable stratification on different triads: the case of $\mathcal{O}(1)$ detuning } \label{Section:5.5} }


{ Previous sections/subsections have revealed that detuning can also be a key factor affecting energy transfer between the waves. Moreover, as shown in figures \ref{fig:mismatchnvariation} and \ref{fig:mismatchnvariation_differnt_triad}, for the same change in stratification and a fixed  parent wave frequency, the detuning introduced in different daughter wave combinations are different.  Therefore, in a medium of varying stratification, it can be expected that triad combinations which undergo less detuning ($\Delta m/m_{\textnormal{min}}$) can exchange more energy in comparison to the triad combinations which are significantly more detuned. The focus of this subsection is to show that, for a given change in background stratification, detuning undergone by different daughter wave combinations of a parent wave, can be a factor in deciding how much energy the daughter waves extract from the parent wave.} 
{ To study this, we also consider $\mathcal{O}(1)$ detuned systems. We remind here that in the previous sections we have considered  $\Delta m/m_{\textnormal{min}} \ll \mathcal{O}(1)$. In order to study the $\mathcal{O}(1)$ (interactions where $\Delta m \sim m_{\textnormal{min}}$) detuned systems, we solve the 2D Boussinesq Navier-Stokes equations using Dedalus.}

{We consider four simulations, where the parent wave has the same angular frequency and horizontal wavenumber. For the same parent wave, two different daughter wave combinations are considered, respectively denoted by `triad $T_{\textnormal{A}}$' and `triad $T_{\textnormal{B}}$'. We choose  the parent wave's angular frequency as $\omega_3/N_b = 0.85$, and following \S \ref{section:5.1.2}, we define a parameter $\alpha$ such that $\omega_1 = (1-\alpha)\omega_3$  and $\omega_2 = \alpha\omega_3$. For `triad $T_{\textnormal{A}}$'  we choose $\alpha = 0.346$, while for `triad $T_{\textnormal{B}}$', $\alpha = 0.312$.}
{ The horizontal wavenumbers of the daughter waves in `triad $T_{\textnormal{A}}$' are $k_1/k_3 = 1.5$ and $k_2/k_3 = -0.5$, and the same for `triad $T_{\textnormal{B}}$' are $k_1/k_3 = 0.6$ and $k_2/k_3 = 0.4$.  The vertical wavenumbers are calculated according to the stratification,  horizontal wavenumbers and angular frequencies. Here we emphasize that `triad $T_{\textnormal{A}}$' belongs to the classification $(k_1/k_3,|{m_1/m_2}|)\in (1,\infty)\times (1,\infty)$, where the detuning is relatively less even for high values of $\omega_3/N_b$. However, `triad $T_{\textnormal{B}}$' belongs to the classification $(k_1/k_3,|{m_1/m_2}|)\in (0,1)\times (0,1)$ where the detuning is significantly higher (shown in figure \ref{fig:mismatchnvariation_differnt_triad}).
We perform four simulations in the following way -- two for `triad $T_{\textnormal{A}}$' in background stratifications $N_b$ and $3N_b$, and likewise for  `triad $T_{\textnormal{B}}$'.
The simulations in base stratification are done using governing equations \eqref{eqn:wave1}--\eqref{eqn:wave3}, since they are resonant at stratification $N_b$. However, the simulations in stratification $3N_b$ are done using Dedalus. }

{ \noindent The initial amplitude profile in the $z$-direction for all the three waves forming the triad is assumed as:
\begin{equation}
a_{j} = ({A}_j/\sqrt{m_{j}}) \ee^{-\left( z/W_{p(j)}\right)^{2}}\ee^{\ii m_{j}z}, 
\label{eqn:amp_def_gauss_O1}
\end{equation}
where $A_j$ is some complex constant such that ${A}_{1}={A}_{2}=6\times10^{-4}$ and ${A}_{3}=2.25\times10^{-2}$ in all four simulations.}
 
{ The width of the wave packets ($W_{p(j)}$) at the base stratification ($N_b$) are chosen to be $W_{p(1)}=W_{p(2)}=W_{p(3)}=44\lambda_3$ for both `triad $T_{\textnormal{A}}$' and `triad $T_{\textnormal{B}}$'.}
 
\begin{figure}
 {\centering
 \includegraphics[width=1.0\linewidth,keepaspectratio]{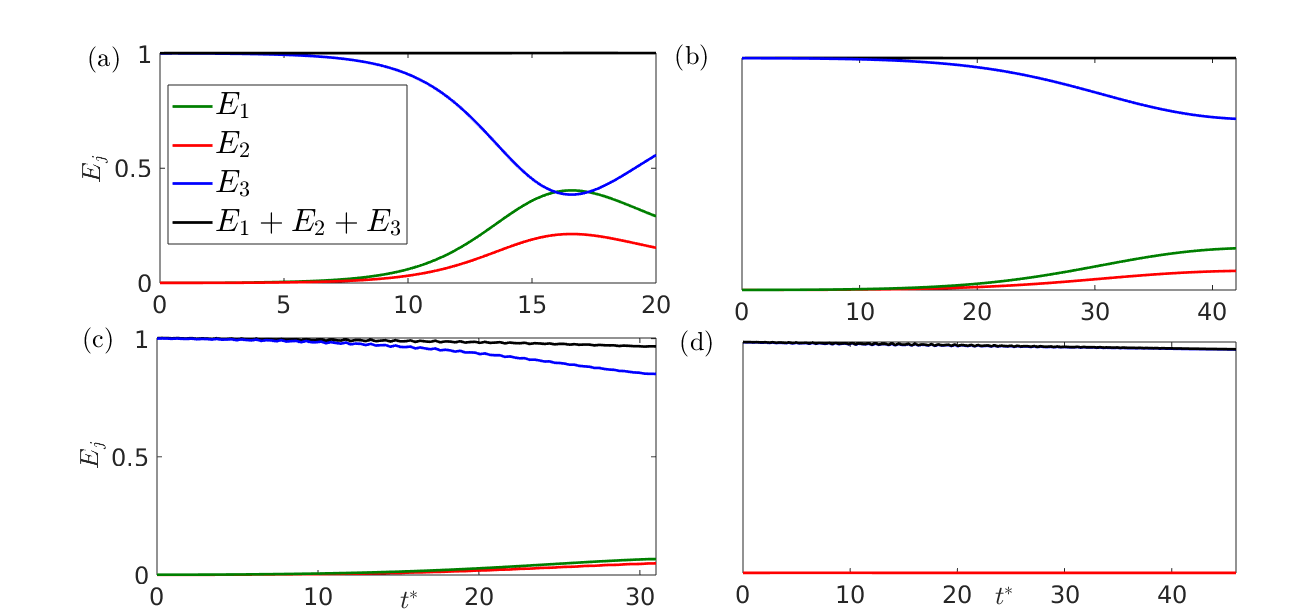}}
 \caption[Energy evolution plot for $\mathcal{O}(1)$ detuned systems.]{{ Comparison of time evolution of energy of `triad $T_{\textnormal{A}}$' and `triad $T_{\textnormal{B}}$' in stratifications $N_b$ and $3N_b$.
  Energy evolution of all waves of  `triad $T_{\textnormal{A}}$' in background stratification of (a) $N_b$ and (c) $3N_b$.   Energy evolution of all waves of  `triad $T_{\textnormal{B}}$' in background stratification of (b) $N_b$ and (d) $3N_b$. Here we again emphasize the point that simulations which has background stratification $N_b$ were run using reduced equations \eqref{eqn:wave1}--\eqref{eqn:wave3}, while simulations which has background stratification $3N_b$ were run using Dedalus.}}
 \label{fig:O1_detuning}
 \end{figure}

{ The widths change as the triads move to a different background stratification (i.e. $3N_b$), which are as follows:}
{ \begin{enumerate}
\item `Triad $T_{\textnormal{A}}$' in stratification $3N_b$: $W_{p(1)}=17.1 \lambda_3 \hspace{0.5cm} W_{p(2)}=15.4\lambda_3 \hspace{0.5cm}  W_{p(3)}=27.4 \lambda_3$.
\item `Triad $T_{\textnormal{B}}$' in stratification $3N_b$:  $W_{p(1)}= 17.6 \lambda_3 \hspace{0.5cm} W_{p(2)}=15.1 \lambda_3 \hspace{0.5cm}  W_{p(3)}=27.4\lambda_3$.
\end{enumerate}
The width and amplitude of the parent wave are chosen such that for both triads $T_{\textnormal{A}}$ and $T_{\textnormal{B}}$, there is a significant exchange of energy at the base stratification ($N_b$) where there is no detuning, see figures \ref{fig:O1_detuning}(a) and \ref{fig:O1_detuning}(b). $T_{\textnormal{A}}$  and $T_{\textnormal{B}}$ undergo different amounts of detuning as they travel to the stratification $3N_b$, which affects their energy transfer differently, see figures \ref{fig:O1_detuning}(c) and \ref{fig:O1_detuning}(d).
While detuning for `triad $T_{\textnormal{A}}$' was small ($\Delta m/m_{\textnormal{min}} = 0.15$), and thus it still exchanged a significant amount of energy (around $18\%$ of the parent waves's energy at $t^*=0$), a reasonably high detuning ($\Delta m/m_{\textnormal{min}} = 0.63$) happened for `triad $T_{\textnormal{B}}$', thereby completely stopping the energy transfer.}

 To summarize, in \S \ref{section:5.1} we analysed the variation of (i) vertical wavenumber detuning, (ii) wave-packets width, (iii) nonlinear coupling coefficients and (iv) group speeds of the wave-packets with stratification. Moreover, we also studied how the  above-mentioned variations effect the energy transfer among the wave-packets under inviscid conditions. In \S \ref{section:5.2}, the effect of viscosity on the growth rates of the daughter waves was studied as the background stratification was varied. It was found that if viscous effects were significant, the growth rates decrease as the background stratification increases. In \S \ref{sec:optimal_base}, it was shown that the optimal base stratification where the parent wave can form a triad without any detuning so that energy transfer would be maximum, is  the highest stratification in the varying medium. In \S \ref{section:5.3}, the results obtained from the reduced order equations were validated with numerical simulations which are done using an open-source pseudo-spectral code Dedalus. 
{In \S \ref{Section:5.5} we considered $\mathcal{O}(1)$ detuned systems, and numerically showed that the sensitivity of vertical wavenumber detuning of a particular daughter wave combination, for a given change in stratification, is also an important factor in deciding how much energy these daughter waves can extract from the parent wave.}
 
\section{Summary and Conclusion  \label{Section:7}} 

To summarize, in this paper we have considered triad interactions among internal gravity waves whose vertical wavelength is at least an order of magnitude smaller than the length scale of buoyancy frequency's vertical variation. 
 Such high wavenumber internal wave triads (or `high modes') significantly influence the energy cascading process that finally leads to ocean turbulence and mixing through PSI. By deriving a simplified, yet fairly generalized  mathematical model, we have studied the energy transfer dynamics in resonant and near-resonant triads in weakly non-uniform stratifications in the presence of viscosity and rotational effects. {To the best of our knowledge, this is the first work where the contribution of each factor (e.g. wave-packet width,  group speed, nonlinear coupling coefficients, detuning in the vertical wavenumber, viscosity) has been delineated in a medium of varying stratification.
 }

First we show that the well-known pump-wave approximation in uniform stratification, although accurately predicts the growth rates of the daughter waves in a detuned triad, it does not give the complete picture of how much energy was actually transferred from the parent wave to the daughter waves. The maximum amount of energy which the parent wave can transfer to the daughter waves is found to be primarily dependent on three factors (which can be combined into a single factor  $\gamma_M/\gamma_A$): (i) group speeds of all three waves,  (ii) the nonlinear coupling coefficients, and (iii) the parent wave's initial amplitude. We emphasize here that the factor $\gamma_M/\gamma_A$ can be evaluated from initial conditions, and can therefore predict the maximum energy transferred during the later stages. 
Therefore, even when normal mode analysis in pump-wave approximation may predict the 
same growth rates for a detuned triad and a resonant triad (having the same horizontal wavenumbers and frequencies), the actual amount of energy transferred in these two cases can be quite different.  Hence pump-wave approximation in near-resonant triads should be used carefully. 

\begin{figure}
{\centering
\includegraphics[width=1.0\linewidth,keepaspectratio]{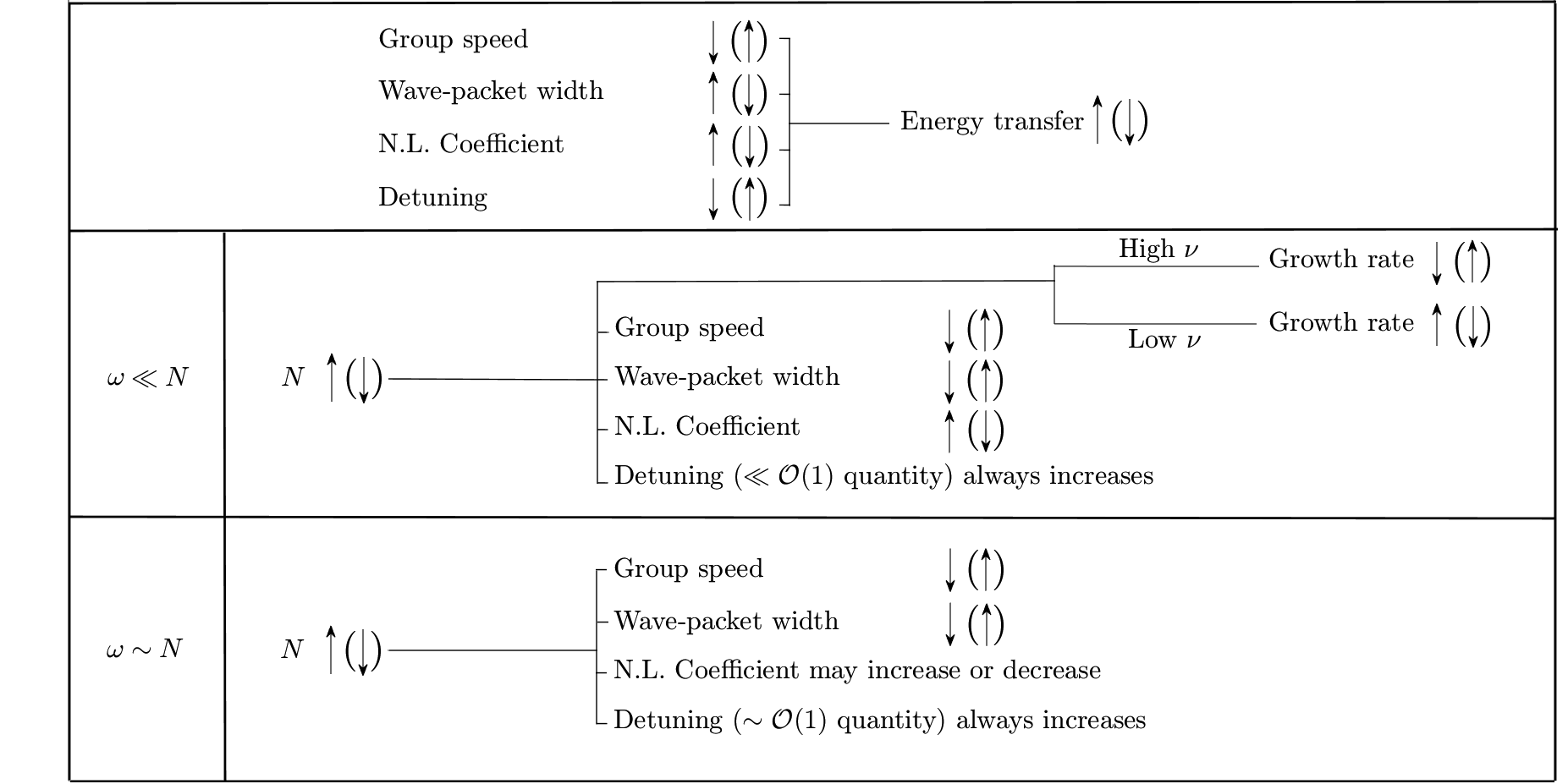}}\caption[Condensed Diagram.]{{ A summary diagram showing how different factors vary as stratification increases ($\uparrow$) or decreases ($\downarrow$). Moreover, it also shows how changing each of these factors affect the energy transfer between the wave packets. }}
\label{fig:Resume_diagram}
\end{figure}

We also consider the interaction between wave-packets forming a near-resonant (i.e. detuned) triad. Two non-dimensional parameters $\Pi_w$ and $\Pi_m$ are defined; $\Pi_w$ depends  mainly on four factors: (i) group speeds of all three waves,  (ii) the nonlinear coupling coefficients, (iii) the parent wave's initial amplitude, and (iv) the width of the wave packets. For $\Pi_m$, although factors (i)--(iii) remain the same as  $\Pi_w$, the fourth factor on which it depends is the detuning in the vertical wavenumber ($\Delta m$).
The mismatch in the vertical wavenumber imposes another constraint on energy transfer between finite width wave-packets. In the parameter regime $\Pi_w \sim \mathcal{O}(1)$ and $\Pi_m \gg \mathcal{O}(1)$, the near-resonant wave-packets need to have a larger width (than its corresponding resonant wave-packets) to exchange the same percentage of energy as the resonant wave-packets. 

Next we considered  energy transfer between wave-packets in weakly varying stratifications under both inviscid and viscous conditions. The main factors which influence the energy transfer in an inviscid scenario are: 
\begin{itemize}
\item  \emph{Group speeds of the wave-packets}: Group speed decreases (increases) when wave-packets travel to a higher (lower) stratification. For a wave satisfying $\omega \ll N$, its group speed is inversely proportional to the local $N$ value. This property is not specific to triads, but any internal wave packet in general.
\item  \emph{Width of the wave-packets}: The width of the wave-packets decreases (increases) when wave-packets travel to a higher (lower) stratification. However, the group speed and the width decreases (increases) by the same factor, hence when the waves interact in a different background stratification, their interaction time remains the same as that in the previous background stratification. Like group speed, this property is not specific to triads, but any internal wave packet in general.
\item  \emph{The nonlinear coupling coefficients}: For triads consisting of waves with frequencies  $\omega_j \ll N$ ($j=1,2,3$), nonlinear coupling coefficients increases (decreases) when waves packets forming a triad travel to a higher (lower) stratification from the base stratification where resonant conditions are perfectly met. The nonlinear coupling coefficients are effectively proportional to the square root of the local stratification value when $\omega_j\ll N$.
For triads with constituent waves satisfying $\omega_j \approx N$,  changes in nonlinear coupling coefficients as stratification changes depend on the daughter waves.
\item \emph{Detuning in the vertical wavenumber}: When wave-packets forming a triad interact in a higher stratification than the base stratification (where resonant condition is perfectly met), depending on the ratio $\omega_3/N$, the effect of detuning can be strong or weak. Triads with parent waves having $\omega_3 \approx N $  can be significantly detuned even for a small changes in the stratification. The daughter waves which satisfy {$(k_1/k_3,|{m_1/m_2}|)\in (1,\infty)\times (0,1)$} undergo the least amount of detuning. Therefore, such triads may be the pathway through which the parent wave decomposes when detuning is highly sensitive to the background stratification.
For a triad with parent waves satisfying $\omega_3\ll N$, the wave detuning is significantly lesser as the stratification is increased, because in this parameter regime, the vertical wavenumber almost behaves as a linear function of the stratification.
\end{itemize}

In a medium of (weakly) varying stratification, the ideal background stratification where the parent wave can form a resonant triad such that the energy transfer is maximum is at the highest stratification. 

Additionally, we also considered \emph{viscous effects} for triads with parent waves satisfying $\omega_3 \ll N$. When viscous effects are significant, the growth rates of the daughter waves decrease (unlike the inviscid case) even when the background stratification is increased from the base stratification (where the resonant condition is perfectly satisfied). This was found to be the case for all possible daughter wave combinations provided the viscosity is high enough.  
{Different daughter wave combinations undergo different amounts of detuning ($\Delta m$) for the same change in the background stratification. By extending the analysis to $\mathcal{O}(1)$ detuned systems it was shown that two different daughter wave combinations (forming a triad with a fixed parent wave), which extract similar amount of energy at the base stratification,  may not extract similar amount of energy in a different background stratification. The daughter waves of the triad which undergo less detuning extract more energy from the parent wave.}

{ The main findings of this study, especially how different factors vary with stratification and their effects on energy transfer, have been briefly summarized in figure \ref{fig:Resume_diagram}.}

\vspace{0.2cm}

\noindent \textbf{Declaration of Interests:} The authors report no conflict of interest.


\section*{Acknowledgement}
The authors are grateful to Subhajit Kar for enlightening discussions  and also helping with the numerical validation. The authors also thank the  anonymous reviewers for useful comments and suggestions.

\appendix
\section{Scaling analysis for finding the relation between the small parameters}
\label{app:A}
Scaling analysis is performed to predict the relation between the time scale of the amplitude's temporal evolution ($\epsilon_{t} t$), length scale of the amplitude function ($\epsilon_{z} z$), the kinematic viscosity $\nu$, and the magnitude of the streamfunction which is taken as an $\mathcal{O}(\epsilon_{a})$ quantity. The length scale of the amplitude function $(a_{j})$ is an input to the system which depends on wave-packets' width, buoyancy frequency profile, detuning in between the triad waves. The parameter $\epsilon_a$ is decided by the amplitude of the waves which is given in the initial conditions.

Let us consider the amplitude evolution equation for a wave-packet (the analysis is similar for all three waves):
\begin{equation}
\ii\left[2\left( \frac{\partial a}{\partial t} - \frac{m(\omega^{2}-f^2)}{\omega(k^{2}+m^{2})} \frac{\partial a}{\partial z}  +  \frac{k(N^{2}-\omega^{2})}{\omega(k^{2}+m^{2})} \frac{\partial a}{\partial x} \right) + \nu\left(k^2+m^2 + \frac{f^2m^2}{\omega^2} \right)a \right]  = \textnormal{RHS}
\label{eqn:orderintro}
\end{equation}


\noindent We neglect the $x$-direction variation in \eqref{eqn:orderintro} for simplicity. In equation \eqref{eqn:orderintro}, the amplitude's evolution with time is assumed to be at least an order lesser than the angular frequency of the wave. Hence the term ${\partial a}/{\partial t}$ will scale as: ${\partial a}/{\partial t} \sim \epsilon_{t} \epsilon_{a} \omega$. In a similar way, amplitude's spatial length scale is assumed to be at least an order less than the vertical wavenumber, therefore the term ${\partial a}/{\partial z}$ will scale as: ${\partial a}/{\partial z} \sim \epsilon_{z} \epsilon_{a} m$.  Hence the LHS of (\ref{eqn:orderintro}) scales as:

\begin{equation}
\ii\epsilon_{a}\left[2\epsilon_t \omega - 2\epsilon_z \frac{m^2(\omega^{2}-f^2)}{\omega(k^{2}+m^{2})} + \nu\left(k^2+m^2 + \frac{f^2m^2}{\omega^2} \right) \right]  = \textnormal{RHS}
\label{eqn:orderLHS}
\end{equation}

The RHS of (\ref{eqn:orderintro}) is given by (we ignore the exponential function since it is an $\mathcal{O}(1)$ quantity):
\begin{equation}
    \textnormal{RHS} =  \mathfrak{N} \epsilon_a^2
    \label{eqn:orderRHS}
\end{equation}{}
In the above equation, the nonlinear coupling coefficient $\mathfrak{N}$ cannot be further simplified. Now comparing LHS and RHS respectively obtained from (\ref{eqn:orderLHS}) and (\ref{eqn:orderRHS}):
\begin{equation}
\left[2\epsilon_t \omega - 2\epsilon_z \frac{m^2(\omega^{2}-f^2)}{\omega(k^{2}+m^{2})} + \nu\left(k^2+m^2 + \frac{f^2m^2}{\omega^2} \right) \right] \sim \mathfrak{N} \epsilon_a
\label{eqn:comp_lhs_rhs}
\end{equation}

The dominant balance can be between any two terms. We mainly focus on three combinations which are given below:
\begin{subequations}
\begin{align}
2\epsilon_t \omega  & \sim \mathfrak{N} \epsilon_a - \left[\nu\left(k^2+m^2 + \frac{f^2m^2}{\omega^2} \right) \right] \label{eqn:scaling_analysis_1}  \\
2\epsilon_t \omega  & \sim \mathfrak{N} \epsilon_a \label{eqn:scaling_analysis_2} \\
2\epsilon_t \omega & \sim \mathfrak{N} \epsilon_a + \left[2\epsilon_z \frac{m^2(\omega^{2}-f^2)}{\omega(k^{2}+m^{2})} \right] \label{eqn:scaling_analysis_3}
\end{align}
\end{subequations}

In (\ref{eqn:scaling_analysis_1}), even though the nonlinear terms and the viscous term can be functions of the spatial coordinate $z$, the equations behave such that $z$ coordinate is a parameter instead of a variable. 
This is because the group speed term is much smaller than the nonlinear and viscous terms. In such kind of systems, detuning in vertical wavenumber will have little or no effect on the growth rates. Similar results were obtained in \cite{pumpcraik}, where all three waves have same group speed.  
In (\ref{eqn:scaling_analysis_2}), the nonlinear term is at least an order of magnitude higher than the viscous and group speed term. This scenario occurs when the group speed and viscosity is small.
In (\ref{eqn:scaling_analysis_3}), the group speed and the nonlinear term influences the energy transfer (for example, see \S \ref{Section:3}, \S \ref{Section:4} and \S \ref{Section:5}). 
An important point to note is that in any particular problem, the approximate value of $\epsilon_{z}$ is decided through the buoyancy frequency profile, wave-packet size or the wave detuning.

\bibliographystyle{jfm} 
\bibliography{jfm-instructions}

\end{document}